\documentclass[11pt,onecolumn]{IEEEtran}
\usepackage[utf8]{inputenc} 
\usepackage[T1]{fontenc}
\usepackage{url}
\usepackage{ifthen}
\usepackage{cite}
\usepackage[cmex10]{amsmath} 
\usepackage{caption}
\usepackage{amsfonts}
\usepackage{amssymb}
\usepackage{color}
\usepackage{graphicx}
\usepackage{algorithmicx}
\usepackage{algpseudocode}
\usepackage{algorithm}

\newtheorem{defn}{Definition}
\newtheorem{thm}{{\cal T}heorem}
\newtheorem{cor}{Corollary}
\newtheorem{prop}{Proposition}
\newtheorem{lem}{Lemma}
\newtheorem{conj}{Conjecture}
\newtheorem{constr}{Construction}
\newtheorem{note}{Remark}
\newtheorem{claim}{Claim}
\newcommand{\bit}{\begin{itemize}}
	\newcommand{\eit}{\end{itemize}}
\newcommand{\bcor}{\begin{cor}}
	\newcommand{\ecor}{\end{cor}}
\newcommand{\beq}{\begin{equation}}
\newcommand{\eeq}{\end{equation}}
\newcommand{\beqn}{\begin{equation}}
\newcommand{\eeqn}{\end{equation}}
\newcommand{\bea}{\begin{eqnarray}}
\newcommand{\eea}{\end{eqnarray}}
\newcommand{\bean}{\begin{eqnarray*}}
	\newcommand{\eean}{\end{eqnarray*}}
\newcommand{\ben}{\begin{enumerate}}
	\newcommand{\een}{\end{enumerate}}
\newcommand{\bdefn}{\begin{defn}}
	\newcommand{\edefn}{\end{defn}}
\newcommand{\bnote}{\begin{note}}
	\newcommand{\enote}{\end{note}}
\newcommand{\bprop}{\begin{prop}}
	\newcommand{\eprop}{\end{prop}}
\newcommand{\blem}{\begin{lem}}
	\newcommand{\elem}{\end{lem}}
\newcommand{\bthm}{\begin{thm}}
	\newcommand{\ethm}{\end{thm}}
\newcommand{\bconj}{\begin{conj}}
	\newcommand{\econj}{\end{conj}}
\newcommand{\bconstr}{\begin{constr}}
	\newcommand{\econstr}{\end{constr}}
\newcommand{\bpf}{\begin{proof}}
	\newcommand{\epf}{\end{proof}}
\newcommand{\bprf}{{\em Proof: }}
\newcommand{\eprf}{\hfill $\Box$}

\begin{document}
	\title{Locally Recoverable Streaming Codes \\
		for Packet-Erasure Recovery} 
	\author{
		\IEEEauthorblockN{Vinayak Ramkumar, Myna Vajha, P. Vijay Kumar \ \\}
		\IEEEauthorblockA{
			Department of Electrical Communication Engineering, IISc Bangalore \\ \{vinram93, mynaramana, pvk1729\}@gmail.com}
		
		\thanks{This  research  is  supported by  the J C Bose National Fellowship JCB/2017/000017.}
	}

	\maketitle
	
	\begin{abstract}
		Streaming codes are a class of packet-level erasure codes that are designed with the goal of ensuring recovery in low-latency fashion, of erased packets over a communication network. It is well-known in the streaming code literature, that diagonally embedding codewords of a $[\tau+1,\tau+1-a]$ Maximum Distance Separable (MDS) code within the packet stream, leads to rate-optimal streaming codes capable of recovering from $a$ arbitrary packet erasures, under a strict decoding delay constraint $\tau$.  Thus MDS codes are geared towards the efficient handling of the worst-case scenario corresponding to the occurrence of $a$ erasures.  In the present paper, we have an increased focus on the efficient handling of the most-frequent erasure patterns.   We study streaming codes which in addition to recovering from $a>1$ arbitrary packet erasures under a decoding delay $\tau$, have the ability to handle the more common occurrence of a single-packet erasure, while incurring smaller delay $r<\tau$. We term these codes as $(a,\tau,r)$ locally recoverable streaming codes (LRSCs), since our single-erasure recovery requirement is similar to the requirement of locality in a coded distributed storage system. We characterize the maximum possible rate of an LRSC by presenting rate-optimal constructions for all possible parameters $\{a,\tau,r\}$. Although the rate-optimal LRSC construction provided in this paper requires large field size, the construction is explicit.   It is also shown that our $(a,\tau=a(r+1)-1,r)$ LRSC construction provides the additional guarantee of recovery from the erasure of $h, 1 \leq h \leq a$, packets, with delay $h(r+1)-1$.  The construction thus offers graceful degradation in decoding delay with increasing number of erasures.
	\end{abstract}
	
	
	\section{Introduction}
	Many next-generation applications such as telesurgery, augmented reality and assisted driving require communication systems with high reliability and low latency. Packet erasures occur in networks due to a variety of reasons and erasure coding is a promising, resource-efficient way, to tackle this problem. Streaming codes are packet-level erasure codes which guarantee packet recovery under a tight decoding deadline. The study of streaming codes was initiated in \cite{MartSunTIT04,MartTrotISIT07}, where packet-level codes capable of recovering each message packet from a burst erasure of size $b$ packets within a decoding delay $\tau$ were studied. A subsequent, more general random and burst erasure sliding window channel model was introduced in \cite{BadrPatilKhistiTIT17}, and streaming codes which ensure packet recovery under decoding delay constraint $\tau$ over this channel model can be found in \cite{BadrRateHalfINFOCOM13, NikPVK,FongKhistiTIT19,NikDeepPVK,NikRamVajKum,DudFongKhi,KhistiExplicitCode,RamVajNikPVK,GSDE,Small}. Many other models of streaming codes have been explored in the literature such as \cite{RudRas,AdlCas,LeongHo,TekHoYaoJag,LeoQurHo}. In the present paper, we focus on channels which introduce erasures at arbitrary locations. For a channel which erases $a>1$ coded packets in the packet stream, we want to ensure that every message packet is recovered within delay $\tau$.  However if only a single coded packet is erased, then message packet recovery should take place under the more stringent delay deadline of $r < \tau$.
	
We use $[a:b]$ to denote $\{a,a+1,\dots,b-1,b\}$. The cardinality of a finite set $S$ is denoted by $|S|$. We denote the linear span of $A \subseteq \mathbb{F}_q^n$ by $\langle A\rangle$. Let $M \in \mathbb{F}_q^{k \times n}$, $I \subseteq [0:k-1]$ and $J \subseteq [0:n-1]$, then $M(I,J)$ is the sub-matrix of $M$ comprising of rows with index in $I$ and columns with index in $J$. If $M$ is a square matrix, then $|M|$  denotes the determinant of $M$.
	
	\subsection{Problem Setup}
	Consider a source with an infinite stream of message packets $\{\underline{m}(t)\}_{t=0}^{\infty}$ which needs to be transmitted to a receiver over an erasure channel. Let $\underline{m}(t)=[m_0(t)~m_1(t)\dots m_{k-1}(t)] \in \mathbb{F}_Q^k$, for all time $t \ge 0$. In order to ensure reliability against packet drops the source first encodes the message packets. In any time slot $t \ge 0$, the source generates and sends coded packet $\underline{c}(t)=[m_0(t)\dots m_{k-1}(t)~p_0(t)\dots p_{n-k-1}(t)] \in \mathbb{F}_Q^n$ and  the receiver receives $\underline{c}(t)$ if it is not erased by the channel. The rate of such a packet-level code is $\frac{k}{n}$. Due to causality of encoder, $\underline{c}(t)$ depends only on present message packet $\underline{m}(t)$ and past message packets $\{\underline{m}(t')\mid t' < t\}$. For $t<0$, we set $\underline{m}(t)=\underline{0}$. An $(a,\tau)$ streaming code (SC) is a packet-level code which guarantees message packet recovery within decoding-delay $\tau$ given that in any sliding window of $(\tau+1)$ packets at most $a$ packet erasures are seen. More formally, for any $t \ge 0$, the message packet $\underline{m}(t)$, can be recovered from packets $ \{\underline{c}(t') \mid t' \in [t:t+\tau] \setminus E\} \cup \{\underline{m}(t') \mid t'<t\} $, for all $E \subseteq [t:t+\tau]$ with $|E| \le a$. Note that $(a,\tau)$ SC exists only if $a \le \tau$. The optimal rate of an $(a,\tau)$ SC is 
	$R_{opt}(a,\tau)=\frac{\tau+1-a}{\tau+1}.$ 
	The rate-optimal $(a,\tau)$ SCs known in the literature \cite{BadrRateHalfINFOCOM13, BadrPatilKhistiTIT17} are obtained by diagonal embedding (DE) of an $[n=\tau+1,k=\tau+1-a]$ MDS code. Here every diagonal in the coded packet stream $(m_0(t),m_1(t+1),\dots,m_{k-1}(t+k-1),p_{0}(t+k),\cdots,p_{n-k-1}(t+n-1))$ is a codeword of the MDS code. We refer readers to Tables~\ref{Tab:R1},~\ref{Tab:R2} for such example constructions. 
	
\begin{figure}
		\vspace{-0.1in}
		\begin{center}
			\includegraphics[width=0.4\textwidth]{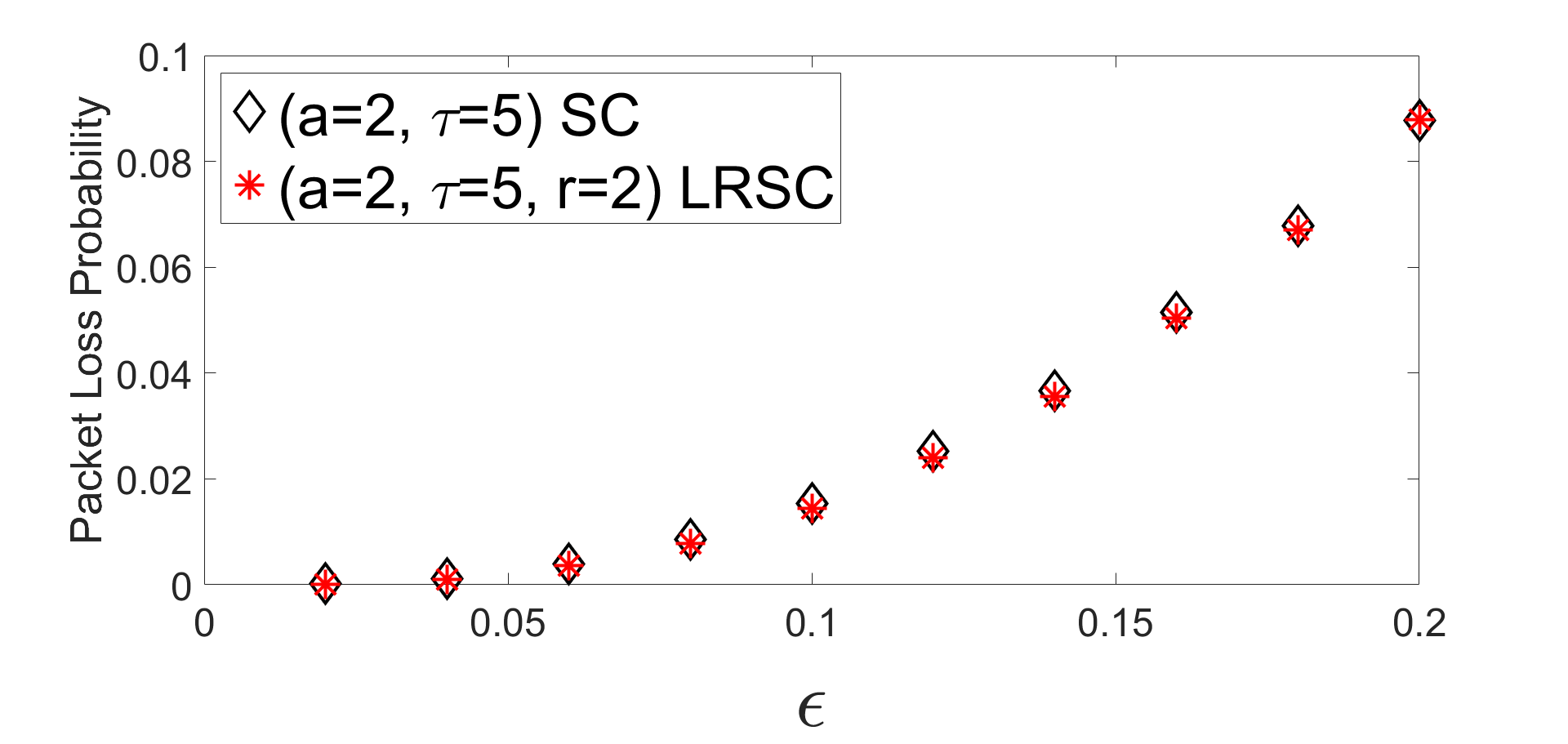}
			\caption{Packet loss probability for $(a=2,\tau=5)$ SC and $(a=2,\tau=5,r=2)$ LRSC over PEC($\epsilon$) channel. }\label{fig:Pe}
		\end{center}
		\vspace{-0.1in}
	\end{figure}

\begin{figure}
		\vspace{-0.1in}
		\begin{center}
			\includegraphics[width=0.4\textwidth]{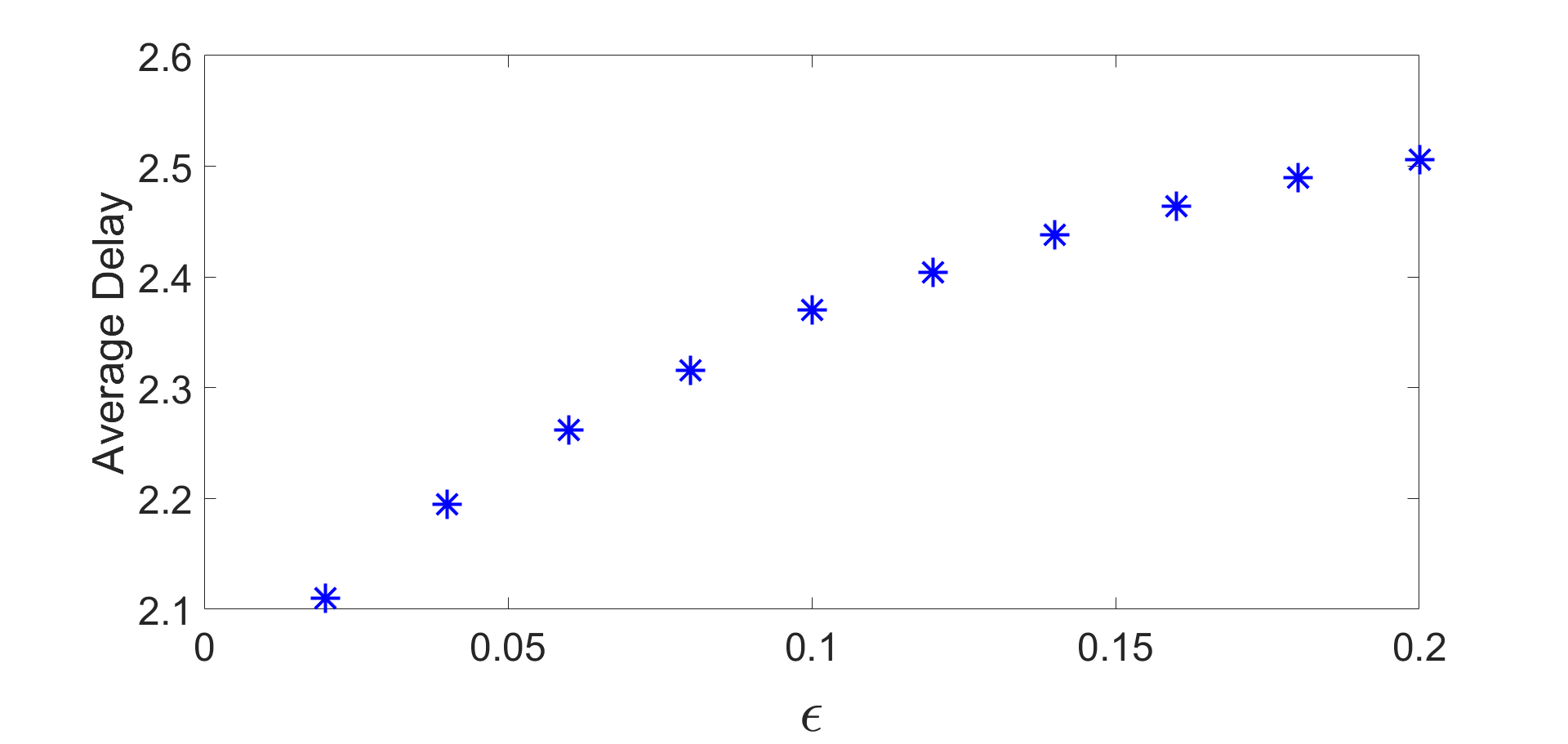}
			\caption{Average delay of recovered packets $(a=2,\tau=5,r=2)$ LRSC for probability over PEC($\epsilon$) channel.} \label{fig:delay}
		\end{center}
		\vspace{-0.25in}
	\end{figure}

	In this work, we study $(a,\tau)$ SCs, for $a>1$, with an additional property that $\underline{m}(t)$, for any $t \ge 0$, should be recoverable from $\{\underline{c}(t') \mid t' \in [t+1:t+r]\} \cup \{\underline{m}(t') \mid t'<t\}$, where $r < \tau$. Such a packet-level code will be referred to as  an $(a,\tau,r)$ locally recoverable streaming code (LRSC). This nomenclature is inspired by the locally recoverable codes in distributed storage literature \cite{HuaCheLi,HanMon,GopHuaSimYek,PapDim,ForYekh,TamBar_LRC,BalNikVajSurvey}, which ensures recovery from single erasure by accessing small number of code symbols. Similarly, if only single coded packet $\underline{c}(t)$ is erased in time window $[t:t+r]$, then an $(a,\tau,r)$ LRSC recovers $\underline{m}(t)$ by time $t+r$, instead of waiting till $t+\tau$. Having a smaller decoding delay for single packet erasure will result in reduced average decoding delay. This is particularly useful for time varying channels with an occasional single packet erasure as the most common event. Consider PEC($\epsilon$) channel where packets get erased randomly and independently with probability $\epsilon$. As shown in Fig.~\ref{fig:Pe}, probability of irrecoverable packet loss over PEC($\epsilon$) channel is almost same for $(a=2,\tau=5,r=2)$ LRSC constructed in this paper and $(a=2,\tau=5)$ SC obtained by DE of $[6,4]$ MDS code. The average delay of recovered packets for $(2,5,2)$ LRSC is much smaller than $\tau=5$ guaranteed by $(2, 5)$ SC, see Fig.~\ref{fig:delay}.  
	
	By definition, an $(a,\tau,r)$ LRSC is a packet-level code which is both an $(a,\tau)$ SC and a $(1,r)$ SC. Hence, the optimal rate of $(a,\tau,r)$ LRSC, denoted by $R_{opt}(a,\tau,r)$, can not exceed $R_{opt}(a,\tau)$ or $R_{opt}(1,r)$, resulting in the following rate upper bound: 
	\bea \label{Eq:bound}
	R_{opt}(a,\tau,r) \le \min\left\{\frac{\tau+1-a}{\tau+1},\frac{r}{r+1}\right\}.
	\eea 
	
	Our problem setup is similar to multicast SCs for two receivers, investigated in \cite{BadrKhiMar_mc1,BadrLuiKhi_mc2, BadrMahKhi_mc3}. Codes capable of handling burst erasure of length $b_1$ with decoding delay $\tau_1$ and $b_2$ length burst erasure with delay $\tau_2$ are studied in  \cite{BadrKhiMar_mc1,BadrLuiKhi_mc2}. The maximum rate of such SCs for almost all  $\{b_1,\tau_1,b_2,\tau_2\}$ parameters are characterized in \cite{BadrLuiKhi_mc2}. In \cite{BadrMahKhi_mc3}, multicast SCs are extended to the case of channels with either same number of arbitrary erasures or burst erasure of different length. We note that these prior works on multicast SCs do not cover the erasure model that we are considering in this paper. 
	\subsection{Our Contributions}
	In this paper, we construct $(a,\tau,r)$ LRSC whose rate matches with the rate upper bound ~\eqref{Eq:bound} for all valid parameters $\{a,\tau,r\}$, leading to our main result given below.
	\bthm
	Let $a$, $\tau$ and $r$ be non-negative integers such that $1 < a \le \tau$ and $r < \tau$. The optimal rate of an $(a,\tau,r)$ LRSC is $R_{opt}(a,\tau,r)=\min\left\{\frac{\tau+1-a}{\tau+1},\frac{r}{r+1}\right\}.$
	\ethm
	
	The rate-optimal LRSCs presented here requires a large field size $q^{2^{(a-2)}}$ where $q \ge r+a-1$, but it is an explicit construction. For all $h \in [1:a]$, the $(a,\tau=a(r+1)-1,r)$ LRSC construction presented in this paper ensures recovery from $h$ packet erasures under delay $h(r+1)-1$.   
	For $(a=2,\tau)$ SCs the previously best-known rate-optimal construction requires a field size $\ge \tau$. Our construction reduces this requirement to $\ge \lceil \frac{\tau+1}{2}\rceil$. 
	
	In Section \ref{Sec:example} we provide an example construction of rate-optimal LRSC. Our rate-optimal LRSC construction for $\tau+1=a(r+1)$ case is presented in Section \ref{Sec:Constr}. This construction is extended to cover all $\{a, \tau, r\}$ parameters in Section \ref{Sec:Ext}.
\section{A Simple Example: $(a=2,\tau=5,r=2)$ LRSC} \label{Sec:example}
Consider DE of any $[n=3,k=2]$ MDS code to obtain a $(1,2)$ SC (see Table~\ref{Tab:R1}). Here, message packet is given by $\underline{m}(t) =[m_0(t)~~m_1(t)]^T$, parity $p_0(t)=m_1(t-2)+m_2(t-1)$ and  coded packet  $\underline{c}(t) =[m_0(t)~~m_1(t)~~p_0(t)]^T$. We will refer to this $(1,2)$  SC as $\mathcal{C}_1$. Now consider DE of an $[n=6,k=4]$ MDS code (see Table~\ref{Tab:R2}). For this case, message packet $\underline{m}(t) =[m_0(t)~~m_1(t)~~m_2(t)~~m_3(t)]^T$ and coded packet $\underline{c}(t) =[m_0(t)~~m_1(t)~~m_2(t)~~m_3(t)~~p_0(t)~~p_1(t)]^T$, where parity symbols $p_0(t) =m_0(t-4)+m_1(t-3)+m_2(t-2)+m_3(t-1)$ and $p_1(t)= m_0(t-5)+ 2m_1(t-4)+ 3m_2(t-3)+ 4m_3(t-2)$ over $\mathbb{F}_5$. This $(2,5)$ SC will be denoted by $\mathcal{C}_2$.
Note that both $\mathcal{C}_1$ and $\mathcal{C}_2$ have rate $\frac{2}{3}$. Also, $R_{opt}(1,2)=R_{opt}(2,5)=\frac{2}{3}$ and hence these codes are rate-optimal SCs. 

\begin{table}[!h]
		\centering
		\caption{DE of $[3,2]_2$ MDS code to yield a rate-optimal $(1,2)$ SC. Here each column represents a coded packet and the symbols colored in red belong to $[3,2]_2$ MDS codeword.}
			\begin{tabular}{ | c | c | c | c | c | c | }
				\hline
				${\color{red} m_0(0)}$ & {\color{blue} $m_0(1)$ }&  $m_0(2)$ & $m_0(3)$ & $m_0(4)$ & $m_0(5)$    \\ \hline
				$m_1(0)$ & {\color{red} $m_1(1)$ }& {\color{blue} $m_1(2)$} &  $m_1(3)$ & $m_1(4)$ & $m_1(5)$    \\ \hline
				-  & $m_1(0)$& {\color{red} \shortstack{ $m_0(0)$ \\ $+m_1(1)$}} & {\color{blue} \shortstack{ $m_0(1)$ \\ $+m_1(2)$}} & \shortstack{ $m_0(2)$ \\ $+m_1(3)$} & \shortstack{ $m_0(3)$ \\ $+m_1(4)$}  \\ \hline
			\end{tabular}
					\label{Tab:R1}
	\end{table}

\begin{table}
	\centering
	\caption{DE of $[6,4]_5$ MDS code to yield a rate-optimal $(2,5)$ SC. Here each column represents a coded packet and symbols in red color correspond to codeword of $[6, 4]_5$ MDS code. \label{Tab:R2}}
		\begin{tabular}{ | c | c | c | c | c | c | c | c |}
			\hline
			{\color{red} $m_0(0)$} & {\color{blue} $m_0(1)$ }&  $m_0(2)$ & $m_0(3)$ & $m_0(4)$ & $m_0(5)$ & $m_0(6)$ & $m_0(7)$ \\ \hline
			$m_1(0)$ & {\color{red} $m_1(1)$ }& {\color{blue} $m_1(2)$} &  $m_1(3)$ & $m_1(4)$ & $m_1(5)$ & $m_1(6)$ & $m_1(7)$  \\ \hline
			$ m_2(0)$ & $m_2(1)$ &  	{\color{red} $m_2(2)$ } & {\color{blue} $m_2(3)$} & $m_2(4)$ & $m_2(5)$ & $m_2(6)$ & $m_2(7)$  \\ \hline
			$m_3(0)$ & $m_3(1)$ &  $m_3(2)$ & 	{\color{red} $m_3(3)$ } & {\color{blue} $m_3(4)$} & $m_3(5)$ & $m_3(6)$ & $m_3(7)$ \\ \hline
			-  & $m_3(0)$&  \shortstack{ $m_2(0)$ \\ $+m_3(1)$} &  \shortstack{ $m_1(0)$ \\ $+m_2(1)$ \\ $+ m_3(2)$} & {\color{red} \shortstack{ $m_0(0)$ \\ $+m_1(1)$ \\ $+m_2(2)$  \\ $+m_3(3)$ }}  & {\color{blue} \shortstack{ $m_0(1)$ \\ $+m_1(2)$ \\ $+m_2(3)$  \\ $+m_3(4)$ }} &    \shortstack{ $m_0(2)$ \\ $+m_1(3)$ \\ $+m_2(4)$  \\ $+m_3(5)$ } &  \shortstack{ $m_0(3)$ \\ $+m_1(4)$ \\ $+m_2(5)$  \\ $+m_3(6)$ } \\ \hline
			-  & - & \shortstack{ $4m_3(0)$} & \shortstack{ $3m_2(0)$ \\ $+4m_3(1)$} & \shortstack{ $2m_1(0)$ \\ $+3m_2(1)$ \\ $+4m_3(2)$} & {\color{red} \shortstack{ $m_0(0)$ \\ $+2m_1(1)$ \\ $+3m_2(2)$  \\ $+4m_3(3)$ } } &  {\color{blue} \shortstack{ $m_0(1)$ \\ $+2m_1(2)$ \\ $+3m_2(3)$  \\ $+4m_3(4)$ }} &  \shortstack{ $m_0(2)$ \\ $+2m_1(3)$ \\ $+3m_2(4)$  \\ $+4m_3(5)$ } \\ \hline
		\end{tabular}
\end{table}
Our aim is to come up with a packet-level code which is both a $(1,2)$ SC and  a $(2,5)$ SC, thus resulting in a $(2,5,2)$ LRSC. We first argue that  $\mathcal{C}_1$ or $\mathcal{C}_2$ can not serve this purpose. Suppose $\mathcal{C}_1$ encoder is employed and assume that coded packets $\underline{c}(0)$ and  $\underline{c}(1)$ are lost. There is no parity symbol in $\{\underline{c}(t) \mid t \in \{2,3,4,5\}\}$ that contains  $m_1(0)$. Hence, the receiver can not recover $m_1(0)$, proving that $\mathcal{C}_1$ is not a $(2, 5)$ SC and hence not an $(2,5,2)$ LRSC. Now imagine that source uses $\mathcal{C}_2$ to encode message packets. If $\underline{c}(0)$ is erased then $m_0(0)$ can not be recovered by accessing only  $\underline{c}(1)$ and $\underline{c}(2)$. Thus, $\mathcal{C}_2$ is not a $(1,2)$ SC and therefore not an $(2,5,2)$ LRSC. 

\begin{table*}
	\begin{center}
			\caption{$(2,5,2)$ LRSC over $\mathbb{F}_3$. Each column represents a coded packet. Parity symbols shown in color red (blue/black) are dependant only on message symbols shown in red (blue/back).}
		\resizebox{.95 \textwidth}{!} 
		{
			\begin{tabular}{ | c | c | c | c | c | c | c | c | c | c | c | }
				\hline
				{\color{red} $m_0(0)$} & {\color{blue} $m_0(1)$ }&  $m_0(2)$ & {\color{red} $m_0(3)$} & {\color{blue} $m_0(4)$ } & $m_0(5)$ & {\color{red} $m_0(6)$ }&  {\color{blue} $m_0(7)$ } & $m_0(8)$  & {\color{red} $m_0(9)$ } & {\color{blue} $m_0(10)$} \\ \hline
				$m_1(0)$ & {\color{red} $m_1(1)$ }& {\color{blue} $m_1(2)$} &  $m_1(3)$ & {\color{red} $m_1(4)$ } & {\color{blue} $m_1(5)$} & $m_1(6)$ & {\color{red} $m_1(7)$ }& {\color{blue} $m_1(8)$ } & $m_1(9)$ & {\color{red}$m_1(10)$}  \\ \hline
				-  & $m_1(0)$& {\color{red} \shortstack{ $m_0(0)$ \\ $+m_1(1)$}} & {\color{blue} \shortstack{ $m_0(1)$ \\ $+m_1(2)$}} & \shortstack{ $2m_1(0)$ \\ $+m_0(2)$ \\ $+m_1(3)$} & {\color{red} \shortstack{ $m_0(0)$ \\ $+2m_1(1)$ \\ $+m_0(3)$ \\ $+m_1(4)$} } & {\color{blue} \shortstack{ $m_0(1)$ \\$+2m_1(2)$ \\ $+m_0(4)$ \\$+m_1(5)$} } &  \shortstack{ $m_0(2)$ \\ $+2m_1(3)$ \\ $+m_0(5)$ \\$+m_1(6)$} &  {\color{red} \shortstack{$m_0(3)$ \\$+2m_1(4)$ \\ $m_0(6)$ \\$+m_1(7)$}} &  {\color{blue} \shortstack{ $m_0(4)$ \\$+2m_1(5)$ \\ $+m_0(7)$ \\$+m_1(8)$ }} & \shortstack{ $m_0(5)$ \\$+2m_1(6)$ \\ $+m_0(8)$ \\$+m_1(9)$ }   \\ \hline
			\end{tabular}
		}
		\label{Tab:R12}	
	\end{center}
\vspace{-0.1in}
\end{table*} 
We now present a $(2,5, 2)$ LRSC, denoted by $\mathcal{C}_{(2,5,2)}$ (see Table~\ref{Tab:R12}). We fix $k=2$ and $n=3$. The message packet is   $\underline{m}(t)=[m_0(t)~~m_1(t)]^T$ and coded packet $\underline{c}(t)=[m_0(t)~~m_1(t)~~p_0(t)]^T$. We construct this code over $\mathbb{F}_3$. The parity symbol $p_0(t)$ is constructed as follows: 
\bean 
p_0(t)=m_0(t-5)+2m_1(t-4)+m_0(t-2)+m_1(t-1).
\eean

Suppose only $\underline{c}(t)$ is erased in $[t:t+2]$ and $\{m(t') \mid t' <t\}$ is known. Then, the receiver obtains $m_0(t)$ from $p_0(t+2)=m_0(t-3)+2m_1(t-2)+{\color{red}m_0(t)}+m_1(t+1)$ and $m_1(t)$ from $p_0(t+1)=m_0(t-4)+2m_1(t-3)+m_0(t-1)+{\color{red}m_1(t)}$. Thus the receiver is able to decode $\underline{m}(t)$ within delay $2$. Therefore, $\mathcal{C}_{(2,5,2)}$ is a $(1, 2)$ SC. Now to show that $\mathcal{C}_{(2,5,2)}$ is a $(2,5)$ SC, consider that packets $\underline{c}(t)$ and $\underline{c}(t+\theta)$ are erased, where $1 \le \theta \le 5$. To show that $\mathcal{C}_{(2,5,2)}$ is a $(2,5,2)$ LRSC, it remains to show that $m_0(t)$ and $m_1(t)$ can be retrieved from $\{\underline{c}(t') \mid t' \in [t+1:t+5] \setminus \{t+\theta\} \}\cup \{\underline{m}(t') \mid t' < t\} $.
 
\paragraph{$\theta=1$} Note that $p_0(t+2)=m_0(t-3)+2m_1(t-2)+{\color{red}m_0(t)+m_1(t+1)}$. The receiver has access to $p_0(t+2)$, $m_0(t-3)$ and  $m_1(t-2)$ as only $\underline{c}(t)$ and $\underline{c}(t+1)$ are erased. Hence  $m_0(t)+m_1(t+1)$ can be obtained. Similarly, using  $p_0(t+5) = {\color{red}m_0(t)+2m_1(t+1)}+m_0(t+3)+m_1(t+4)$, $m_0(t+3)$ and  $m_1(t+4)$ it is possible to get $m_0(t)+2m_1(t+1)$. Using these two, $m_0(t)$ can be recovered.
Now for decoding of $m_1(t)$, use $p_0(t+4)=m_0(t-1)+2{\color{red}m_1(t)}+m_0(t+2)+m_1(t+3)$, in which all other symbols are known.
\paragraph{$\theta=2$} Here $m_0(t)$ can be obtained from $p_0(t+5)={\color{red}m_0(t)}+2m_1(t+1)+m_0(t+3)+m_1(t+4)$ since only $\underline{c}(t)$ and $\underline{c}(t+2)$ are unknown.
The decoding of $m_1(t)$ is carried out utilizing $p_0(t+1)=m_0(t-4)+2m_1(t-3)+m_0(t-1)+{\color{red}m_1(t)}$.
\paragraph{$\theta=\{3,4,5\}$} Since we have shown that $\mathcal{C}_{(2,5,2)}$ is a $(1,2)$ SC. $m_0(t)$ and $m_1(t)$ can be recovered from $\underline{c}(t+1)$ and $\underline{c}(t+2)$. 
 
Thus $\mathcal{C}_{(2,5,2)}$ can handle single packet erasure within delay $2$ and two packet erasures within delay $5$. Hence $\mathcal{C}_{(2,5,2)}$ is an $(2,5,2)$ LRSC of rate $\frac{2}{3}$. From the upper bound in \eqref{Eq:bound} we have $R_{opt}(2,5,2) \le \frac{2}{3}$, which leads to $R_{opt}(2,5,2) = \frac{2}{3}$. 
\begin{note} The rate-optimal $(2,5)$ SC known in the literature is obtained by DE of $[6,4]$ MDS code, which needs a field of size $\ge 5$. The $\mathcal{C}_{(2,5,2)}$ code presented here is also a rate-optimal $(2,5)$ SC and it requires only a smaller size field $\mathbb{F}_3$. 
\end{note}

\section{LRSC Construction for $\tau+1=a(r+1)$} \label{Sec:Constr}
In this section we will describe construction of a rate-optimal  $(a,\tau,r)$ LRSC for the case $\tau+1=a(r+1)$. We will later show in next section how to relax this condition to give optimal rate constructions for all possible parameters. Note that when $\tau+1=a(r+1)$, we have $R_{opt}{(a,\tau)}=\frac{\tau+1-a}{\tau+1}=\frac{r}{r+1}=R_{opt}(1,r)$. Thus the upper bound \eqref{Eq:bound} becomes 
$R_{opt}(a,a(r+1)-1,r) \le \frac{r}{r+1}$. In this section we will show the construction of an $(a,a(r+1)-1,r)$ LRSC of rate $\frac{r}{r+1}$, thereby proving achievability for this case. 
We will refer to this rate $\frac{r}{r+1}$ packet-level code as $\mathcal{C}_{(a,a(r+1)-1,r)}$. 

The LRSC construction has $k=r$, $n=r+1$ and $Q=q^{2^{(a-2)}}$ where $q \ge r+a-1$ be a prime power. 
The message packet $\underline{m}(t)=[m_0(t)~\dots~m_{r-1}(t)]^T \in \mathbb{F}_Q^r$ and coded packet $\underline{c}(t)=[m_0(t)~\dots~m_{r-1}(t)~p_0(t)]^T \in \mathbb{F}_Q^{r+1}$, where $p_0(t)\in \mathbb{F}_Q$ is a parity symbol. Therefore defining $p_0(t)$ completes definition of the LRSC construction.

We will now introduce some notation to define the LRSC construction. We assume $m_i(t)=0$ for $t<0$ and define $(1 \times r)$ diagonal message vector $\hat{m}(t)$ for all $t \ge 0$ as:
\bea \label{Eq:diag}
\underline{\hat{m}}(t)=\left[m_0(t)~m_1(t+1)~\dots~m_{r-1}(t+r-1)\right].
\eea 
	Let $C \in \mathbb{F}_q^{r \times a}$ be an $(r \times a)$ matrix  such that every square sub-matrix of $C$ is non-singular. It is possible to pick such a matrix for $q \ge r+a-1$.  This can be argued as follows. Let $\tilde{G}=[I_{r}~\tilde{P}]$ be the generator matrix of an $[n=r+a,k=r]$ MDS code in systematic form. Then every square sub-matrix of $\tilde{P}$ is non-singular and we can choose $C=\tilde{P}$.   Note that $[n,k]$ MDS codes over $\mathbb{F}_q$ are known for  $q \ge n-1$ and doubly extended Reed-Solomon code \cite{macwilliamssloane} is an MDS code with $q=n-1$  . 
	
 Let $Q_0=Q_1=q$ and $Q_j = q^{2^{(j-1)}}$ for  $j \in [2:a-1]$.  Note that $Q_{a-1}=Q=q^{2^{(a-2)}}$. Suppose $\alpha_0=\alpha_1=1$ and $\alpha_j \in \mathbb{F}_{Q_{j}} \setminus \mathbb{F}_{Q_{j-1}}$ for  $j \in [2:a-1]$. Now form an $(a \times a)$ diagonal matrix $A=diag(\alpha_0, \dots, \alpha_{a-1}) \in \mathbb{F}_{Q}^{a \times a}$ and obtain $(r \times a)$ matrix 
\bea \label{Eq:Gamma_defn}
\Gamma=CA = [\underline{\Gamma}_0~\underline{\Gamma}_1~\cdots~\underline{\Gamma}_{a-1}] \in \mathbb{F}_{Q}^{r \times a}.
\eea
\begin{constr}
	With the above notation, the parity symbol $p_0(t)$ of $\mathcal{C}_{(a,a(r+1)-1,r)}$ for all $t \ge 0$ is defined as follows:
	\bea \label{Eq:parity}
	p_0(t)=\sum_{j=0}^{a-1}\underline{\hat{m}}\left(t-r-j(r+1)\right) \underline{\Gamma}_{j}.
	\eea 
\end{constr}
\subsection*{Example:$(a=3,\tau=8,r=2)$ LRSC}
For this example, $k=2$, $n=3$, $q=4$ and $Q=16$. Suppose $C=\left[ \begin{array}{ccc}
c_{0,0} & c_{0,1} & c_{0,2}   \\
c_{1,0} & c_{1,1} & c_{1,2}   \\
\end{array}\right]$  be a $(2 \times 3)$ matrix over $\mathbb{F}_4$ whose every square sub-matrix is non-singular. 
Let $\alpha_2=\alpha \in \mathbb{F}_{16} \setminus \mathbb{F}_4$ and hence $A=\left[ \begin{array}{ccc}
1 & 0 & 0   \\
0 & 1 & 0   \\
0 & 0 & \alpha 
\end{array}\right]$. Then,
$$\Gamma=CA=\left[ \begin{array}{ccc}
c_{0,0} & c_{0,1} & \alpha c_{0,2}   \\
c_{1,0} & c_{1,1} & \alpha c_{1,2}   \\
\end{array}\right]=\left[\underline{\Gamma}_0~\underline{\Gamma}_1~\underline{\Gamma}_2\right].$$
The parity symbol $p_0(t)$ of rate $\frac{2}{3}$ packet-level code $\mathcal{C}_{(3,8,2)}$  over $\mathbb{F}_{16}$ has the form 
\bean 
p_0(t)&=& \underline{\hat{m}}(t-2)\underline{\Gamma}_0 + \underline{\hat{m}}(t-5)\underline{\Gamma}_1 + \underline{\hat{m}}(t-8)\underline{\Gamma}_2\\
&=&c_{0,0}m_0(t-2)+c_{1,0}m_1(t-1)+c_{0,1}m_0(t-5)+\\ && c_{1,1}m_1(t-4)+\alpha c_{0,2}m_0(t-8)+ \alpha c_{1,2}m_1(t-7),
\eean 
as shown in Table~\ref{Tab:R13}.

\begin{table*}
	\begin{center}
			\caption{$(3,8,2)$ LRSC over $\mathbb{F}_{16}$. Each column represents a coded packet.}
		\resizebox{.99 \textwidth}{!} 
		{
			\begin{tabular}{ | c | c | c | c | c | c | c | c | c | c | c | c | c |}
				\hline
				{\color{red} $m_0(0)$} & {\color{blue} $m_0(1)$ }&  $m_0(2)$ & {\color{red} $m_0(3)$} & {\color{blue} $m_0(4)$ } & $m_0(5)$ & {\color{red} $m_0(6)$ }&  {\color{blue} $m_0(7)$ } & $m_0(8)$  & {\color{red} $m_0(9)$ } & {\color{blue} $m_0(10)$} & $m_0(11)$  & {\color{red} $m_0(12)$}\\ \hline
				$m_1(0)$ & {\color{red} $m_1(1)$ }& {\color{blue} $m_1(2)$} &  $m_1(3)$ & {\color{red} $m_1(4)$ } & {\color{blue} $m_1(5)$} & $m_1(6)$ & {\color{red} $m_1(7)$ }& {\color{blue} $m_1(8)$ } & $m_1(9)$ & {\color{red} $m_1(10)$}  & {\color{blue} $m_1(11)$ } & $m_1(12)$\\ \hline
				-  & $c_{1,0}m_1(0)$ & {\color{red} \shortstack{ $c_{0,0}m_0(0)$ \\ $+c_{1,0}m_1(1)$}} & {\color{blue} \shortstack{ $c_{0,0}m_0(1)$ \\ $+c_{1,0}m_1(2)$}} & \shortstack{ $c_{1,1}m_1(0)$ \\ $+c_{0,0}m_0(2)$ \\ $+c_{1,0}m_1(3)$} & {\color{red} \shortstack{ $c_{0,1}m_0(0)$ \\ $+c_{1,1}m_1(1)$ \\ $+c_{0,0}m_0(3)$ \\ $+c_{1,0}m_1(4)$}} & {\color{blue} \shortstack{$c_{0,1}m_0(1)$ \\ $+c_{1,1}m_1(2)$ \\ $+c_{0,0}m_0(4)$ \\ $+c_{1,0}m_1(5)$}} & \shortstack{  $\alpha c_{1,2}m_1(0)$ \\ $+c_{0,1}m_0(2)$ \\ $+c_{1,1}m_1(3)$ \\ $+c_{0,0}m_0(5)$ \\ $+c_{1,0}m_1(6)$} &  {\color{red} \shortstack{$\alpha c_{0,2}m_0(0)$ \\ $+\alpha c_{1,2}m_1(1)$ \\ $+c_{0,1}m_0(3)$ \\ $+c_{1,1}m_1(4)$ \\ $+c_{0,0}m_0(6)$ \\ $+c_{1,0}m_1(7)$}} & {\color{blue} \shortstack{$\alpha c_{0,2}m_0(1)$ \\ $+\alpha c_{1,2}m_1(2)$ \\ $+c_{0,1}m_0(4)$ \\ $+c_{1,1}m_1(5)$ \\ $+c_{0,0}m_0(7)$ \\ $+c_{1,0}m_1(8)$}}  &  \shortstack{$\alpha c_{0,2}m_0(2)$ \\ $+\alpha c_{1,2}m_1(3)$ \\ $+c_{0,1}m_0(5)$ \\ $+c_{1,1}m_1(6)$ \\ $+c_{0,0}m_0(8)$ \\ $+c_{1,0}m_1(9)$} &  {\color{red} \shortstack{$\alpha c_{0,2}m_0(3)$ \\ $+\alpha c_{1,2}m_1(4)$ \\ $+c_{0,1}m_0(6)$ \\ $+c_{1,1}m_1(7)$ \\ $+c_{0,0}m_0(9)$ \\ $+c_{1,0}m_1(10)$}} & {\color{blue} \shortstack{$\alpha c_{0,2}m_0(4)$ \\ $+\alpha c_{1,2}m_1(5)$ \\ $+c_{0,1}m_0(7)$ \\ $+c_{1,1}m_1(8)$ \\ $+c_{0,0}m_0(10)$ \\ $+c_{1,0}m_1(11)$}}  \\ \hline
			\end{tabular}
		}
		\label{Tab:R13}
	\end{center}
\vspace{-0.1in}
\end{table*}

We will first show that this is a $(1, r=2)$ SC. If $\underline{c}(t)$ is erased and $\{\underline{c}(t+1),\underline{c}(t+2)\} \cup \{\underline{m}(t') \mid t' <t\}$ are known, then the receiver can recover $m_0(t)$ from $p_0(t+2)$ and $m_1(t)$ from $p_0(t+1)$. Therefore $\mathcal{C}_{(3,8,2)}$ is a $(1,r=2)$ SC.

We will now show that the construction results in an $(a=3, \tau=8)$ SC. Suppose $a=3$ coded packets $\underline{c}(t)$, $\underline{c}(t+\theta_1)$ and $\underline{c}(t+\theta_2)$ are erased, where $\theta_1,\theta_2 \in [1:8]$, and $\{m(t') \mid t' <t\}$ is known. Then the receiver needs to decode $\underline{m}(t)$ by time $t+8$ for $\mathcal{C}_{(3,8,2)}$ to be a $(3,8)$ SC. 
Let $\mathcal{C}^*$ be a $[9,6]$ code over $\mathbb{F}_{16}$ with parity check matrix $$H=\left[P^T~~I_3\right],$$ where $$P^T = \left[\begin{array}{ccc}
\underline{\Gamma}_0^T & 0 & 0\\
\underline{\Gamma}_1^T & \underline{\Gamma}_0^T  & 0 \\
\underline{\Gamma}_2^T & \underline{\Gamma}_1^T  & \underline{\Gamma}_0^T 
\end{array}\right].$$ 
Note the definition of $\underline{\hat{m}}(t)$ in equation~\eqref{Eq:diag}. Let us define:
\bea
\label{Eq:reducedpc}[\hat{p}_0(t+2) \ \hat{p}_0(t+5) \ \hat{p}_0(t+8)] = [\underline{\hat{m}}(t) \ \underline{\hat{m}}(t+3) \ \underline{\hat{m}}(t+6)]P.
\eea
 Thus $\hat{p}_0(t+2)=\underline{\hat{m}}(t)\underline{\Gamma}_0$ and $\hat{p}_0(t+5)=\underline{\hat{m}}(t)\underline{\Gamma}_1 +\underline{\hat{m}}(t+3)\underline{\Gamma}_0$.
Note that $\hat{p}_0(t+2)$ and $\hat{p}_0(t+5)$ can be obtained from $p_0(t+2)$ and $p_0(t+5)$ respectively, by removing contribution of message symbols before time $t$ in equation \eqref{Eq:parity}. 

\begin{claim} \label{claim:cstarprop}
If for $\underline{c}=(c_0,c_1,c_2,c_3,c_4,c_5,c_6,c_7,c_8) \in \mathcal{C}^*$, $i \in \{0,1\}$ and $\theta_1,\theta_2 \in [i+1:8]$, $c_i$ is recoverable from $\{c_j \mid j \in [i+1:8]\setminus \{\theta_1,\theta_2
\}\} \cup \{c_j \mid j \in [0:i-1]\} $, then $\mathcal{C}_{(3,8,2)}$  is an $(a=3,\tau=8)$ SC. 
\end{claim}
\bprf
From \eqref{Eq:reducedpc} and definition of $\underline{\hat{m}}(t)$, $(m_0(t), m_1(t+1),m_0(t+3),m_1(t+4),m_0(t+6),m_1(t+7),\hat{p}_0(t+2),\hat{p}_0(t+5), p_0(t+8))$ is a codeword of $\mathcal{C}^*$. Therefore, $m_0(t)$  can be recovered from any $6$ symbols in $\{m_1(t+1),m_0(t+3),m_1(t+4),m_0(t+6),m_1(t+7),\hat{p}_0(t+2),\hat{p}_0(t+5), p_0(t+8)\}$. $6$ symbols in this set of $8$ symbols should be available as there are only $a-1 = 2$ more packet erasures in $[t+1:t+8]$.

Similarly, $(m_0(t-1), m_1(t), m_0(t+2), m_1(t+3),m_0(t+5),m_1(t+6),\hat{p}_0(t+1),\hat{p}_0(t+4), \hat{p}_0(t+7))$ is a codeword in ${\cal C}^*$ from equation \eqref{Eq:reducedpc} by setting $t=t-1$. Therefore, $m_1(t)$ can be obtained using $m_0(t-1)$ and any $5$ symbols from $\{m_0(t+2),m_1(t+3),m_0(t+5),m_1(t+6),\hat{p}_0(t+1),\hat{p}_0(t+4), p_0(t+7)\}$. $5$ symbols in this set of $7$ symbols should be available as there are only $a-1 = 2$ more packet erasures in $[t+1:t+8]$.  Thus, the ${\cal C}_{3, 8, 2}$ is an $(3, 8)$ SC if $\mathcal{C}^*$ satisfies the above mentioned property. 
\eprf

%

We will now show that ${\cal C}^*$ defined using the parity check matrix $H=[P^T \ I_3]$ given by:
\bean 
H=\left[ \begin{array}{cc|cc|cc|ccc}
	c_{0,0} & 	c_{1,0} & 0 & 0 & 0 & 0 & 1 & 0 & 0 \\ c_{0,1} &  c_{1,1} & c_{0,0} & c_{1,0} & 0 & 0 & 0 & 1 & 0 \\ \alpha c_{0,2}  & \alpha c_{1,2}   & c_{0,1}  & c_{1,1} & c_{0,0}  & c_{1,0} & 0 & 0 & 1
\end{array}\right].  
\eean
satisfies the properties mentioned in Claim\ref{claim:cstarprop}. It then follows by Claim\ref{claim:cstarprop} that ${\cal C}_{3, 8, 2}$ is a $(3, 8)$ SC. Let $\underline{h}_j$ denote the $j$-th column of $H$, for $j \in [0:8]$. To show that $c_i$ is recoverable from $\{c_j \mid j \in [i+1:8]\setminus \{\theta_1,\theta_2
\}\} \cup \{c_j \mid j \in [0:i-1]\} $ it is enough to show that $\underline{h}_i$ is not in the span of $\{\underline{h}_{\theta_1},\underline{h}_{\theta_2}\}$. 

\paragraph{$i=0$} If $\{\theta_1,\theta_2\} \cap \{1,6\}=\phi$, then clearly $\underline{h}_0 \notin \left<\{\underline{h}_{\theta_1},\underline{h}_{\theta_2}\}\right>$ as $H(0,j)=0$  for $j \in [1:8]\setminus \{1,6\}$ and $H(0,0)=c_{0,0} \ne 0$. Suppose $\{\theta_1,\theta_2\} = \{1,6\}$, then 
$|H([0:2],\{0,1,6\})|=\alpha|C([0:1],[1:2])| \ne 0$. Therefore, $\underline{h}_0 \notin \left<\{\underline{h}_{1},\underline{h}_{6}\}\right>$.  Now let $\theta_1=1$ and $\theta_2 \in \{2,3,4,5,7,8\}$. It can be verified that $H{[0:2],\{0,1,\theta_2\}})$ is invertible for $\theta_2 \in \{4,5,7,8\}$ due to the invertibility of square sub-matrices of $C$. 
For $\theta_2 \in \{2,3\}$, $H{[0:2],\{0,1,\theta_2\}})$ is invertible since $|H{[0:2],\{0,1,\theta_2\}})|=c_{\theta_2-2,0}|C([0:1],[0:1])|+c_{\theta_2-2,1}\alpha |C([0:1],\{0,2\}| \ne 0$.
Now consider the case  $\theta_1=6$ and $\theta_2 \in \{2,3,4,5,7,8\}$. From $H(j,6)=0$ for $j \in [1:2]$, it can be argued that $\underline{h}_0 \in \left<\{\underline{h}_{6},\underline{h}_{\theta_2}\}\right>$ only if there exists $\beta \in \mathbb{F}_{16}$ such that $\beta H(\theta_2,1)=c_{0,1}$ and $\beta H(\theta_2,2)=\alpha c_{0,2}$. Since $H(\theta_2,1),c_{0,1},H(\theta_2,2),c_{0,2} \in \mathbb{F}_4$ and $\alpha \in \mathbb{F}_{16} \setminus \mathbb{F}_4$, such a $\beta$ does not exist. Hence, $\underline{h}_0 \notin \left<\{\underline{h}_{6},\underline{h}_{\theta_2}\}\right>$.

\paragraph{$i=1$} Note that $H(0,j)=0$  for $j \in [2:8]\setminus \{6\}$ and hence $\underline{h}_1 \notin \left<\{\underline{h}_{\theta_1},\underline{h}_{\theta_2}\}\right>$ if $6 \notin \{\theta_1,\theta_2\}$. Suppose $\theta_1=6$ and $\theta_2 \in \{2,3,4,5,7,8\}$. Using arguments similar that used above, it can be easily seen that there is no $\beta \in  \mathbb{F}_{16}$ such that $\beta H(\theta_2,1)=c_{1,1}$ and $\beta H(\theta_2,2)=\alpha c_{1,2}$. Therefore, $\underline{h}_1 \notin \left<\{\underline{h}_{6},\underline{h}_{\theta_2}\}\right>$. Thus we have argued that  $\underline{h}_i \notin \left<\{\underline{h}_{\theta_1},\underline{h}_{\theta_2}\}\right>$ for any  $i \in \{0,1\}$ and $\theta_1,\theta_2 \in [i+1:8]$, thereby proving that $\mathcal{C}_{(3,8,2)}$ is a $(3,8,2)$ LRSC. \\

It can be shown that $\mathcal{C}_{(3,8,2)}$ is also a $(2,5)$ SC. 
Consider any $\underline{c}=(c_0,c_1,c_2,c_3,c_4,c_5,c_6,c_7,c_8) \in \mathcal{C}^*$. By Claim 1, $c_i$ is recoverable from $\{c_j \mid j \in [i+1:8]\setminus \{\theta,8\}\} \cup \{c_j \mid j \in [0:i-1]\} $, for any $i \in \{0,1\}$ and $\theta \in [i+1:7]$. From the structure of $\mathcal{C}^*$, observe that message symbols $c_4$ and $c_5$ have no contribution to parity symbols $c_6$ and $c_7$. Hence, $c_i$ is recoverable from $\{c_j \mid j \in [i+1:8]\setminus \{\theta, 4,5,8 \}\} \cup \{c_j \mid j \in [0:i-1]\} $. Using arguments similar to that used in the proof of Claim 1, it can be shown that $m_0(t)$ can be obtained from any $4$ symbols in $\{m_1(t+1),m_0(t+3),m_1(t+4),\hat{p}_0(t+2),\hat{p}_0(t+5)\}$ and $m_1(t)$ can be obtained using $m_0(t-1)$ and any $3$ symbols from $\{m_0(t+2),m_1(t+3),\hat{p}_0(t+1),\hat{p}_0(t+4) \}$. Thus, if coded packets $\underline{c}(t)$ and $\underline{c}(t_1)$ are erased, where $t_1 \in [t+1:t+5]$, we can recover $\underline{m}(t)$ by time $t+5$.

\begin{table*}
	\begin{center}
			\caption{$(2,4,2)$ LRSC over $\mathbb{F}_3$. Each column represents a coded packet.}
		\resizebox{.95 \textwidth}{!} 
		{
			\begin{tabular}{ | c | c | c | c | c | c | c | c | c | c | c | }
				\hline
				{\color{red} $m_0(0)$} & {\color{blue} $m_0(1)$ }&  $m_0(2)$ & $m_0(3)$ & $m_0(4)$  & {\color{red} $m_0(5)$ } & {\color{blue} $m_0(6)$} &  $m_0(7)$  & $m_0(8)$  &  $m_0(9)$  & {\color{red}$m_0(10)$} \\ \hline
				$m_1(0)$ & {\color{red} $m_1(1)$ }& {\color{blue} $m_1(2)$} &  $m_1(3)$ &  $m_1(4)$  & $m_1(5)$ & {\color{red} $m_1(6)$ } & {\color{blue} $m_1(7)$} &  $m_1(8)$  & $m_1(9)$ & $m_1(10)$  \\ \hline
				$m_2(0)$ &  $m_2(1)$ &  $m_2(2)$ & {\color{red} $m_2(3)$ }&  {\color{blue}$m_2(4)$}  &  $m_2(5)$ & $m_2(6)$ &  $m_2(7)$ & {\color{red}$m_2(8)$}  & {\color{blue}$m_2(9)$} & $m_2(10)$  \\ \hline
				-  & $m_1(0)$ & {\color{red} \shortstack{ $m_0(0)$ \\ $+m_1(1)$}} & {\color{blue} \shortstack{ $m_0(1)$ \\ $+m_1(2)$}} & \shortstack{ $m_0(2)$ \\ $+m_1(3)$ \\ $+m_2(0)$} & \shortstack{ $m_0(3)$ \\ $+m_1(4)$ \\ $+m_2(1)$ } & \shortstack{ $m_0(4)$ \\ $+m_1(5)$ \\ $+m_2(2)$} & {\color{red} \shortstack{  $m_0(5)$ \\ $+m_1(6)$ \\ $+m_2(3)$}} & {\color{blue} \shortstack{ $m_0(6)$ \\ $+m_1(7)$ \\ $+m_2(4)$ }} & \shortstack{ $m_0(7)$ \\ $+m_1(8)$ \\ $+m_2(5)$} & \shortstack{ $m_0(8)$ \\ $+m_1(9)$ \\ $+m_2(6)$}  \\ \hline
				-  & $m_2(0)$ & $m_2(1)$ & \shortstack{ $2m_1(0)$ \\ $+m_2(2)$ } & {\color{red} \shortstack{ $m_0(0)$ \\ $+2m_1(1)$ \\ $+m_2(3)$ } } & {\color{blue} \shortstack{ $m_0(1)$ \\$+2m_1(2)$ \\ $+m_2(4)$ } }  &  \shortstack{ $m_0(2)$ \\ $+2m_1(3)$ \\ $+m_2(5)$ } & \shortstack{ $m_0(3)$ \\ $+2m_1(4)$ \\ $+m_2(6)$ } & \shortstack{ $m_0(4)$ \\ $+2m_1(5)$ \\ $+m_2(7)$ }& {\color{red} \shortstack{ $m_0(5)$ \\ $+2m_1(6)$ \\ $+m_2(8)$ }}&  {\color{blue} \shortstack{ $m_0(6)$ \\ $+2m_1(7)$ \\ $+m_2(9)$ }}\\ \hline
			\end{tabular}
		}
		\label{Tab:R12_}
	\end{center}
\vspace{-0.1in}
\end{table*}

\subsection{Proof that $\mathcal{C}_{(a,a(r+1)-1,r)}$ is an $(a,a(r+1)-1,r)$ LRSC}
Assume that a single coded packet $\underline{c}(t)$ is erased in time window $[t:t+r]$ and that all the message packets before $t$ are known. Pick any $i \in [0:r-1]$, $m_i(t)$ is an element in $\underline{\hat{m}}(t-i)$. By the definition of parity check in equation \eqref{Eq:parity},
all symbols involved in $p_0(t+r-i)$, other than $m_i(t)$, are known. Hence, the receiver can decode $m_i(t)$ using $p_0(t+r-i)$ for any $i \in [0:r-1]$ and thus $\underline{m}(t)$ is recoverable within delay $r$. We have thus argued that $\mathcal{C}_{(a,a(r+1)-1,r)}$ is a $(1,r)$ SC. In order to prove that $\mathcal{C}_{(a,a(r+1)-1,r)}$ is a rate-optimal $(a,a(r+1)-1,r)$ LRSC we need to show that it is also an $(a,a(r+1)-1)$ SC. We first reduce this proof to showing certain code symbol recovery properties for a scalar code, as stated in the Lemma~\ref{Lem:scalar} below. Let $P \in \mathbb{F}_Q^{ar \times a}$ be an $(ar \times a)$ matrix defined as follows:
\bea
\label{Eq:P_defn}
P^T = \left[\begin{array}{cccccc}
	\underline{\Gamma}_0^T & 0 & 0 & \cdots & 0 & 0\\
	\underline{\Gamma}_1^T & \underline{\Gamma}_0^T & 0 & \cdots& 0 & 0\\
	\vdots & \vdots & \ddots & \vdots & \vdots\\
	 \underline{\Gamma}_{a-2}^T &  \underline{\Gamma}_{a-3}^T & \cdots & \cdots& \underline{\Gamma}_0^T & 0\\
	\underline{\Gamma}_{a-1}^T & \underline{\Gamma}_{a-2}^T & \cdots & \cdots & \underline{\Gamma}_1^T  & \underline{\Gamma}_0^T \\
\end{array} \right]
\eea
where $\Gamma=[\underline{\Gamma}_0~~\dots~~\underline{\Gamma}_{a-1}]$ is the $(r \times a)$ matrix defined in \eqref{Eq:Gamma_defn}. 

\blem \label{Lem:scalar} 
Let $\mathcal{C}^*_{a,r}$ be a $[a(r+1),ar]$ scalar code over $\mathbb{F}_{Q}$ with parity check matrix $H=\left[P^T \ \ -I_a\right]$. 
If for all codewords $\underline{c}=(c_0,c_1,\dots,c_{a(r+1)-1}) \in \mathcal{C}^*_{a,r}$ and erasure sets ${\cal E} \subseteq [0:a(r+1)-1]$ with $|{\cal E}|=a$, $\{c_i \mid i \in {\cal E} \cap [0:r-1] \}$ is recoverable from unerased code symbols $\{c_j \mid j \in [0:a(r+1)-1]\setminus {\cal E} \}$, then $\mathcal{C}_{(a,a(r+1)-1,r)}$  is an $(a,\tau)$ SC.
\elem 
\bprf
To show that $\mathcal{C}_{(a,a(r+1)-1,r)}$  is an $(a,\tau)$ SC, it is enough to show that for any $t \ge 0$, $\underline{m}(t)$ can be recovered from $\{ c(t') \mid t' \in [t:t+a(r+1)-1] \setminus E \} \cup \{c(t') \mid t' < t\}$  where $t \in E$ and $|E| = a$. In order to recover $m_i(t)$ which is an element in $\underline{\hat{m}}(t-i)$, let us consider $a$ parity checks $p_0(t-i+r), p_0(t-i+2r+1), \cdots, p_0(t-i+a(r+1)-1)$ in which $\underline{\hat{m}}(t-i)$ participates. From equation \eqref{Eq:parity} we have:
\bean
p_0(t-i+\ell(r+1)-1) = \sum\limits_{j=0}^{a-1} \underline{\hat{m}}(t-i+(\ell-j-1)(r+1)) \Gamma_j.
\eean
Note that $\underline{\hat{m}}(t')$ is known for all $t' < t-r+1$ as we know all the message symbols before time $t$. Therefore we can obtain $\hat{p}_0(t-i+\ell (r+1)-1)$ from $p_0(t-i+\ell (r+1)-1)$ where
%
%
\bean
\hat{p}_0(t-i+\ell (r+1)-1)=\sum_{j=0}^{\ell-1}\underline{\hat{m}}\left(t-i-(\ell-j-1)(r+1)\right) \underline{\Gamma}_{j}.
\eean
Set $\hat{p}(t-i)=[\hat{p}_0(t-i+r)~\hat{p}_0(t-i+2(r+1)-1)~\dots~\hat{p}_0(t-i+a(r+1)-1)]$. Then, it follows that $\underline{c}^{(i)}=(\hat{m}(t-i),\hat{m}(t-i+r+1),\dots,\hat{m}(t-i+(a-1)(r+1)+1),\hat{p}(t-i))$ is a codeword of $\mathcal{C}^*_{a,r}$ and $m_i(t)$ is $i$-th symbol of codeword $\underline{c}^i$. Note that the codeword $\underline{c}^{(i)}$ contains $a(r+1)$ symbols from $a(r+1)$ packets with index in $[t-i:t-i+a(r+1)-1]$ and  $a$ packet erasures in $[t:t+a(r+1)]$ imply at most $a$ erasures in codeword $\underline{c}^{(i)}$. The recovery property of $\mathcal{C}^*_{a,r}$ guarantees that $m_i(t)$ can be obtained from unerased symbols in $\underline{c}^{(i)}$. Thus $\underline{m}(t)$ is recoverable within delay $a(r+1)-1$. 
\eprf

We prove that $\mathcal{C}^*_{a,r}$ satisfies the recovery properties stated in Lemma~\ref{Lem:scalar} using a parity check viewpoint. The following result connects code symbol recovery with properties of parity check matrix for any scalar linear code.  
\blem  \label{Lem:Recovery}
Let $\mathcal{C}$ be an $[n,k]$ scalar code and $H=[\underline{h}_0~\underline{h}_1~\dots~\underline{h}_{n-1}] \in \mathbb{F}_Q^{(n-k) \times n}$ be a parity check matrix of $\mathcal{C}$. Suppose the code symbols indexed by coordinates in set $\mathcal{E} \subseteq [0:n-1]$ are erased. Let $m \le n$ be a positive integer and $\mathcal{I} = \mathcal{E} \cap [0:m-1]$. Then, for any codeword $\underline{c}=(c_0, c_1,\dots, c_{n-1}) \in \mathcal{C}$ we have the following result.
\bit 
\item $\{c_j \mid j \in \mathcal{I}\}$ can be recovered from unerased code symbols $\{ c_j \mid j \in [0:n-1] \setminus \mathcal{E} \}$  if  $$\underline{h}_{i} \notin  \left<\{\underline{h}_{j} \mid j \in \mathcal{E} \cap [i+1:n-1] \}\right>$$ for all $i \in \mathcal{I}$.
\eit 
\elem 
\bprf
The Lemma follows directly from a well-known result, nevertheless we provide a brief proof for it. Suppose $\underline{h}_{i} \notin  \left<\{\underline{h}_{j} \mid j \in \mathcal{E} \cap [i+1:n-1] \}\right>$ for all $i \in \mathcal{I}$. Assume that there exists a codeword   $\underline{c}=(c_0,c_1,\dots,c_{n-1}) \in \mathcal{C}$ such that $\{c_j \mid j \in \mathcal{I}\}$ is not recoverable from unerased code symbols $\{ c_j \mid j \in [0:n-1] \setminus \mathcal{E} \}$. Let $i_0$ be the smallest integer in $\mathcal{I}$ such that $c_{i_0}$ is not recoverable from $\{ c_j \mid j \in [0:n-1] \setminus \mathcal{E} \} \cup \{ c_j \mid j \in [0:i_0-1] \}$. For this to happen, there should exist another codeword $\underline{d}=(d_0,d_1,\dots,d_{n-1}) \in \mathcal{C}$ such that $d_{i_0} \ne c_{i_0}$ and $d_j = c_j ~\forall j \in ([0:n-1]\setminus \mathcal{E})\cup [0:i_0-1]$. Since $H$ is parity check matrix of $\mathcal{C}$ and $\underline{c},\underline{d} \in \mathcal{C}$ we have,
\bean 
\sum\limits_{j=0}^{n-1}c_j\underline{h}_j=\sum\limits_{j=0}^{n-1}d_j\underline{h}_j=\underline{0} \implies \sum\limits_{j=0}^{n-1}(c_j-d_j)\underline{h}_j=\underline{0}.
\eean 
As $d_j = c_j ~\forall j \in ([0:n-1]\setminus \mathcal{E})\cup [0:i_0-1]$ we get, 
\bea \label{Eq:Recovery} 
(c_{i_0}-d_{i_0})\underline{h}_{i_0}+\sum\limits_{j \in \mathcal{E} \cap [i_0+1:n-1]}(c_j-d_j)\underline{h}_j=\underline{0}.
\eea 
Since $c_{i_0} \ne d_{i_0}$, it follows from \eqref{Eq:Recovery} that  $\underline{h}_{i_0} \in \left<\{\underline{h}_{j} \mid j \in \mathcal{E} \cap [i_0+1:n-1] \}\right>$, which results in a contradiction. Therefore, there no such $i_0 \in \mathcal{I}$ and no such unrecoverable codeword $\underline{c}\in \mathcal{C}$. 
\eprf 

Before moving to that proof that $\mathcal{C}^*_{a,r}$ meets the recovery conditions stated in Lemma \ref{Lem:scalar}, we first prove some results on $\Gamma$ and $P$ matrices which are useful for the proof.  
\bdefn (Interference matrix) An $(r \times a)$ matrix $D=(d_{i,j}) \in \mathbb{F}_{Q_{a-2}}^{r \times a}$ will be referred to as an interference matrix  if $d_{i,j}=0$ if $j \in \{0,1\}$ and $d_{i,j} \in \mathbb{F}_{Q_{j-1}}$ for $j \in [2:a-1]$. 
\edefn 
We note that $D=\mathbf{0}_{r \times a}$ is an example of interference matrix.
\blem \label{Lem:inv0} Let $D \in \mathbb{F}_{Q_{a-2}}^{r \times a}$ be an interference matrix and $\Gamma \in \mathbb{F}_{Q}^{r \times a}$ be the matrix defined in \eqref{Eq:Gamma_defn}. Then, any square sub-matrix of  $\Gamma+D$ is non-singular.
\elem
\bprf Pick any two sets $I \subseteq [0:r-1]$  and $J \subseteq [0:a-1]$ of same cardinality $z$. In order to prove the lemma we need to show that $$U=\Gamma(I,J)+D(I,J)=C(I,J)A(J,J)+D(I,J)$$ is non-singular. Let $\hat{C} = C(I,J)$,  $\hat{A} = A(J,J)$ and $\hat{D} = D(I,J)$. Thus $U=\hat{C}\hat{A}+\hat{D}$. Let $J =\{j_0,j_1,\dots,j_{z-1}\}$ with $j_0<j_1<\dots <j_{z-1}$. Then, $\hat{A}=diag(\alpha_{j_0}, \alpha_{j_1},\dots,\alpha_{j_{z-1}})$. Since $\hat{C}$ is a square sub-matrix of $C$, by definition $|\hat{C}| \neq 0$.  
We define $(z \times z)$ matrices
\bean 
U^{(t)}=\left[U([0:r-1],[0:t-1])~~\hat{C}([0:r-1],[t:z-1])\right]
\eean
for $t \in [1:z-1]$, $U^{(0)}=\hat{C}$ and $U^{(z)}=U$. We will now show by induction that $U^{(t+1)}$ is invertible given $U^{(t)}$ is invertible. 
Clearly $U^{(0)} = \hat{C}$ is invertible.
\bean
U^{(t+1)} &=& \left[U([0:r-1],[0:t])~~\hat{C}([0:r-1],[t+1:z-1])\right] \\
&=& \left[U([0:r-1],[0:t-1]) \ \ \ \ \alpha_{j_t} \hat{C}([0:r-1],t)+ \hat{D}([0:r-1],t) \ \ \  \hat{C}([0:r-1],[t+1:z-1])\right]
\eean
Let us define matrices,
\bean 
W^{(t)}= \left[U([0:r-1],[0:t-1]) \ \ \ \  \hat{D}([0:r-1],t) \ \ \  \hat{C}([0:r-1],[t+1:z-1])\right]
\eean
for $t \in [0:z]$. Now it is clear to see that for $t \in [0:z-1]$:
\bean 
|U^{(t+1)}|= \alpha_{j_{t}}|U^{(t)}|+|W^{(t)}|.
\eean 


If $j_{t+1} \in \{0,1\}$ 
we have $U^{(t+1)}= \hat{C}$ and hence is non-singular. Now consider $j_{t+1} > 1$. Then, $\alpha_{j_{t+1}} \in \mathbb{F}_{Q_{j_{t+1}}} \setminus  \mathbb{F}_{Q_{j_{t}}}$ and $|W^{(t)}|,|U^{(t)}| \in \mathbb{F}_{Q_{j_{t}}} \subset \mathbb{F}_{Q_{j_{t+1}}}$. Therefore we have $|U^{(t+1)}| \in \mathbb{F}_{Q_{j_{t+1}}} \setminus \mathbb{F}_{Q_{j_{t}}}$ if  $|U^{(t)}| \neq 0$. Hence,  $|U^{(t+1)}| \neq 0$ given  $|U^{(t)}| \neq 0$. Now since $|U^{(0)}|= |\hat{C}| \neq 0$, by repeated application of this result we have $|U|=|U^{(z)}| \neq 0$, proving that $U=\Gamma(I,J)+D(I,J)$ is non-singular.
\eprf 

\begin{figure}
	\begin{center} 
		\includegraphics[width=0.5\textwidth]{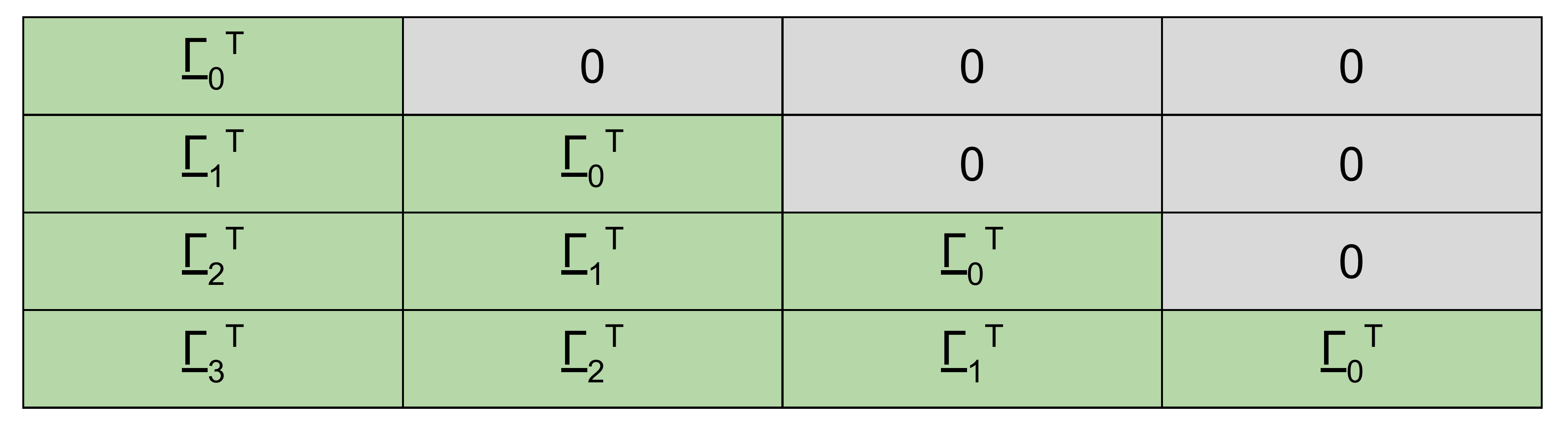}		
		\caption{Structure of $4 \times 4r$ matrix $\mathcal{P}^T$ for $a=4$.}
		\label{Fig:rho}
	\end{center}
\end{figure}

Now we look at the parity check matrix  $H = \left[\mathcal{P}^T~-I_{a}\right] \in \mathbb{F}_Q^{a \times a(r+1)}$ of $\mathcal{C}^*_{a,r}$, see Fig.~\ref{Fig:rho}. The Lemma~\ref{Lem:proof} given below in conjunction with Lemma~\ref{Lem:Recovery} proves that  $\mathcal{C}^*_{a,r}$ has the required recovery properties for  $\mathcal{C}_{(a,a(r+1)-1,r)}$ to be an $(a,\tau)$ SC.
\blem \label{Lem:proof}
Consider parity check matrix  $H = \left[\mathcal{P}^T~-I_{a}\right] =[\underline{h}_0~ \underline{h}_1~\dots~\underline{h}_{a(r+1)-1}] \in \mathbb{F}_Q^{a \times a(r+1)}$ of $\mathcal{C}^*_{a,r}$. Let $\mathcal{E} \subseteq [0:a(r+1)-1]$ be an erasure set with $|{\cal E}|=a$ and $|{\cal E} \cap [0:r-1]|>0$. Then, $\underline{h}_{i} \notin  \left<\{\underline{h}_{j} \mid j \in \mathcal{E} \cap [i+1:a(r+1)-1] \}\right>$ for all $i \in \mathcal{E} \cap [0:r-1]$.
\elem 
\bprf 
We first divide the erasure set $\mathcal{E}$ in to different segments. Define $\mathcal{E}_i=\mathcal{E} \cap [ir:ir+r-1]$ and $e_i=|\mathcal{E}_i|$  for all $i \in [0:a-1]$. Note that $e_0>0$ by definition of $\mathcal{E}$. We also set $\mathcal{\hat{E}}=\mathcal{E} \cap [ar:a(r+1)-1]$ and $\hat{e}=|\mathcal{\hat{E}}|$. Now we look at the unerased parity symbols. Let $f_i=\big{|}[ar:ar+i] \setminus \mathcal{E}\big{|}$ for all $i \in [0:a-1]$. Then, we have $$\hat{e}+f_{a-1}=a=|\mathcal{E}|=\sum_{i=0}^{a-1}e_i+\hat{e}$$ and hence $\sum_{i=0}^{a-1}e_i=f_{a-1}$. This means that the number of message symbols erased is same the number of unerased parity symbols. Now pick the smallest integer  $\ell \in [0:a-1]$ such that $\sum_{i=0}^{\ell}e_i=f_{\ell}$. It follows from above arguments that such an $\ell \in [0:a-1]$ always exists. 

First we consider $\mathcal{E}$ such that $\ell=0$. Since $e_0>0$ and $f_0 \le 1$ by definition, $\ell=0$ occurs only if $e_0=f_0=1$. Let $i$ be this single element in $\mathcal{E} \cap [0:r-1]$, then from structure of $H$ (see equation~\eqref{Eq:P_defn}), it follows that $H(0,j)=0$ for $j \in \mathcal{E} \setminus \{i\}$ and $H(0,i) \ne 0$. Therefore, the Lemma holds for $\ell=0$. From now onwards we  consider $\ell>0$ and hence $e_0>f_0$. 

Let us consider non-erased parities with index $\le \ell$ given by $$X=[0:\ell] \setminus (\{j-ar\mid j \in \mathcal{\hat{E}}\})$$ and erased symbols with index $\le \ell r+r-1$ $$Y= \mathcal{E} \cap [0:\ell r+r-1].$$ We note that $|X|=f_\ell=\sum_{i=0}^{\ell}e_i=|Y|$.
Now define $(f_\ell \times f_\ell)$ sub-matrix $\hat{H}=H(X,Y)$ and let $\underline{\hat{h}}_j$ denote $j$-th column of $\hat{H}$ for all $j \in [0:f_\ell-1]$. By the definition shown in equation~\eqref{Eq:P_defn}, $\hat{H}$ has the following structure:
\bean
\scalebox{0.9}{$
\left[ \begin{array}{ccccc}
\Gamma( {\cal E}_0', B_0)^T & 0 & 0 & \cdots & 0 \\
\Gamma({\cal E}_0', B_1)^T & \Gamma( {\cal E}_1', B_1-1)^T & 0 & \cdots & 0 \\
\vdots & \vdots & \ddots & \ddots & \vdots\\
\Gamma( {\cal E}_0', B_{\ell})^T & \Gamma({\cal E}_1', B_{\ell}-1)^T & \cdots & \cdots & \Gamma( {\cal E}_{\ell}', B_{\ell}-\ell)^T\\
\end{array} \right]$}
\eean
where ${\cal E}_i' = \{ j-ai \mid j \in {\cal E}_i\}$ and $B_i = \{i\} \setminus \{j-ar \mid j \in \hat{\cal E}\}$.
It can be seen that $H(i,j)=0$ if $i \in X$ and $j \in \mathcal{E} \setminus Y$ from equation \eqref{Eq:P_defn}. Hence, if $\underline{\hat{h}}_{p} \notin  \left<\{\underline{\hat{h}}_{j} \mid j \in  [i+1:f_\ell-1] \}\right>$ for all $p \in [0:e_0-1]$ then $\underline{h}_{i} \notin  \left<\{\underline{h}_{j} \mid j \in \mathcal{E} \cap [p+1:a(r+1)-1] \}\right>$ for all $i \in \mathcal{E} \cap [0:r-1]$. Suppose $e_0=f_{\ell}$, then 
\bean
\hat{H} = H(X, {\cal E}_0) = \Gamma({\cal E}_0, X)^T \text{ by equation \eqref{Eq:P_defn}}.
\eean
As any $(e_0 \times e_0)$ sub-matrix of $\Gamma$ is invertible, $\hat{H}$ is invertible and hence Lemma is true for this case. We now consider the case where $f_{\ell}>e_0>f_0$ case. It follows for this case that $\sum_{i=0}^{j}e_{i}>f_{j}$ for all $j \in [0,\ell-1]$. Suppose $\sum_{i=0}^{j}e_{i} = f_{j}$ contradicts definition of $\ell$. If  $\sum_{i=0}^{j}e_{i} < f_{j}$ it implies that, $\sum_{i=0}^{j-1}e_{i} < f_{j-1}$ for $e_j=1$ and if $e_j = 0$ then $\sum_{i=0}^{j-1}e_{i} < f_{j-1} + \mathbf{1}_{ \{ar+j \in {\cal E} \}}$ implying $\sum_{i=0}^{j-1}e_{i} \le f_{j-1}$. Applying this repeatedly contradicts that $e_0 > f_0$.

Let $S=[\underline{s}_0~\underline{s}_1\dots~\underline{s}_{f_\ell-1}]$ be an $(f_\ell \times f_\ell)$ matrix obtained from $\hat{H}$ by applying elementary row operations. Then,   
$\underline{s}_{p} \notin  \left<\{\underline{s}_{j} \mid j \in  [i+1:f_\ell-1] \}\right>$ implies $\underline{\hat{h}}_{p} \notin  \left<\{\underline{\hat{h}}_{j} \mid j \in  [i+1:f_\ell-1] \}\right>$. Thus, in order to prove the lemma it is sufficient to come up with an $(f_\ell \times f_\ell)$ matrix $S$ such that
\bit
\item $S$ is obtainable from  $\hat{H}$ through elementary row operations and
\item there exists a subset $A \subseteq [0:f_{\ell}-1]$ with $|A|=e_0$ such that $S(A,[0:e_0-1])$ is non-singular and $S(A,[e_0:f_{\ell}-1])=\mathbf{0}_{e_0 \times f_{\ell}-e_0}$.
\eit  
We obtain this $(f_{\ell} \times f_{\ell})$ matrix $S$ and $e_0$ element set $A$ using Algorithm~\ref{algo}. 
\begin{algorithm}[H]
	Input: $\hat{H}$,$e_0,\dots,e_{\ell}$, $f_0,\dots,f_{\ell}$ \\
	Output: $S$, $A$ \\
	$i \gets \ell$, $A \gets [0:f_{\ell}-1]$, $S \gets \hat{H}$
	\begin{algorithmic}[1]
		\While {$i>0$}
		\If {$e_i=!0$}
		\State $A_i \gets A \cap [f_{i-1}:f_{\ell}-1]$ 
		\State $\hat{A}_i \gets$ smallest $e_i$ elements of $A_i$ 
		\State	$C_i \gets \left[\sum_{u=0}^{i-1}e_j:\sum_{u=0}^{i}e_j-1\right]$ 
		\State Add linear combinations of rows in $\hat{A}_i$  to rows in 
		$A_i \setminus \hat{A}_i$ of $S$  so that  $S(A_i \setminus \hat{A}_i,C_i) \gets \mathbf{0}$
		\State	$A \gets A \setminus \hat{A}_i$
		\EndIf
		\State	$i \gets i-1$
		\EndWhile
	\end{algorithmic}
	\caption{Row reduction}
	\label{algo}
\end{algorithm}
Now we argue the correctness of Algorithm~\ref{algo}. Let $p \in [0:\ell-1]$ such that, $e_p \ne 0$ and $e_i = 0$ for all $i > p$. This implies that $\sum\limits_{i=0}^p e_i = f_{\ell}$ and that $e_p < f_{\ell} - f_{p-1}$ as we know that $\sum\limits_{i=0}^{p-1} e_i > f_{p-1}$. Therefore in step 4 of the algorithm, we can pick $e_p$ smallest elements from $f_{\ell}-f_{p-1}$ elements with $A_p=[f_{p-1}:f_{\ell}-1]$. By the structure of the matrix $\hat{H}$:
\bean
S(A_p, C_i) = \Gamma({\cal E}_p, D)^T,
\eean
where, $D = \left\lbrace j-ar-p \mid j \in [ar+p:ar+\ell] \setminus \hat{\cal E} \right\rbrace$. It can be verified that $|D| = f_{\ell}-f_{p-1}$.
Any $(e_p \times e_p)$ submatrix of $S(A_p, C_i)$ is invertible by the cauchy property of $\Gamma$. Therefore we can row reduce to generate $S(A_p \setminus \hat{A}_p, C_i) = 0$.

Fix some $j \in [1:\ell]$ with $e_j \ne 0$ and assume that steps $i=\ell,\ell-1,\dots,j+1$ of algorithm are over. The $i=j$ iteration will go through if $|A_j| \ge e_j$ and $S(\hat{A_j},C_j)$ is non-singular at the beginning of step 6.
In each step $i$ the size  of $A$ reduces by $e_i$ and hence 
\bean
|A_j| &=& f_{\ell}-f_{j-1}-\sum_{u=j+1}^{\ell}e_{u}\\
&=& e_j + \sum_{u=0}^{j-1} e_u - f_{j-1}\\
&>& e_j \text{ as } \sum_{u=0}^{j-1} e_u > f_{j-1}.
\eean
Step 6 of the algorithm goes through, if $S(A_j, C_j)$ is invertible. We will show that this is true for any $j \in [1:\ell]$ such that $e_j \ne 0$. Let $x \in \hat{A}_j$ and $y \in C_j$. By definition, either $\hat{H}(x,y) \in \mathbb{F}_q$ or $\hat{H}(x,y)=\alpha c$, where $c \in \mathbb{F}_q$ and $\alpha \in \mathbb{F}_{Q_{\theta}} \setminus \mathbb{F}_{Q_{\theta-1}}$ for some $\theta \in [1:a-1]$. Let $\delta_u$ be the element added to $(x,y)-$th entry due to row reductions carried out in step 6 of algorithm for $u \in [j+1:\ell]$ such that $e_u > 0$. If $\hat{H}(x,y) \in \mathbb{F}_q$, then $\hat{H}(x,y) = \underline{\Gamma}_0(y-\sum\limits_{i=0}^j e_i)$ or $\hat{H}(x,y) = \underline{\Gamma}_1(y-\sum\limits_{i=0}^j e_i)$. In both these cases it can be verified there will not be any row reductions performed on this row.




\subsubsection*{Claim 1} If $\hat{H}(x,y)=\alpha c$ for $x \in \hat{A}_i$ and $y \in C_i$, where $c \in \mathbb{F}_q$ and $\alpha \in \mathbb{F}_{Q_{\theta}} \setminus \mathbb{F}_{Q_{\theta-1}}$, then $\delta_u \in  \mathbb{F}_{Q_{\theta-1}}$ where $\delta_u$ is a component added to $\hat{H}(x,y)$ during step $u\in[i+1:\ell]$. 

\bprf 
Clearly this statement is true for $i = \ell$ as $\delta_{\ell} = 0 \in \mathbb{F}_{Q_{\theta-1}} $. Now let us assume that is true for all $j >= i+1$. We will show that it is true for $j = i$. 

Suppose at step $u \in [j+1:\ell]$, row reduction is applied on row $x \in \hat{A}_j$. It implies that $x \notin \hat{A}_u$ and the row reduction is done to cancel out $S(x, C_u)$ to $0$. By the structure of $\hat{H}$, $S(x, C_u) \in \mathbb{F}_{Q_{\theta-1}} $ when $\hat{H}(x,y) \in \mathbb{F}_{Q_{\theta}}$. Therefore $\delta_u \in \mathbb{F}_{Q_{\theta-1}}$.
\eprf 

At the beginning of step $j$, from the Claim 1 we have that every entry in $S(\hat{A_j},C_j)$ either belongs to $\mathbb{F}_q$ or has the form $\alpha c+ \delta$, with  $\alpha \in \mathbb{F}_{Q_{\theta}} \setminus \mathbb{F}_{Q_{\theta-1}}$ and $\delta \in \mathbb{F}_{Q_{\theta-1}}$. It can be seen that Lemma~\ref{Lem:inv0} is applicable here and hence $S(\hat{A}_j,C_j)$ is invertible. Thus step $j$ goes through.

At the end of $i=1$ step, we get $|A|=f_{\ell}-\sum_{u=1}^{\ell}e_{u}=e_0$, $S(A,[\sum_{i=0}^{i-1}e_{1}:f_{\ell}-1])=\mathbf{0}$ and $S(A,[0:e_0-1])$ non-singular.
\eprf 

We now show an additional property of our $(a,a(r+1)-1,r)$ LRSC construction that it can recovery form any $h \in [1:a]$ erasures within delay $h(r+1)-1$.
\begin{lem}
	For any  $h \in [1:a]$, $\mathcal{C}_{(a,a(r+1)-1,r)}$ is an $(h,h(r+1)-1)$ SC.
\end{lem}	
\bprf 
If for all  $\underline{c}=(c_0,c_1,\dots,c_{a(r+1)-1}) \in \mathcal{C}^*_{a,r}$ and ${\cal E} \subseteq [0:hr-1] \cup [ar:ar+h-1]$ with $|{\cal E}|=h$, $\big\{c_i \mid i \in {\cal E} \cap [0:r-1] \big\}$ can be obtained from $\big\{c_j \mid j \in \big([0:hr-1]\cup [ar:ar+h-1]\big) \setminus {\cal E} \big\}$, then $\mathcal{C}_{(a,a(r+1)-1,r)}$  is an $(h,h(r+1)-1)$ SC. This relation between $\mathcal{C}_{(a,a(r+1)-1,r)}$ and $\mathcal{C}^*_{a,r}$ follows from arguments similar to that used in the proof of Lemma~\ref{Lem:scalar}. Pick any ${\cal E} \subseteq [0:hr-1] \cup [ar:ar+h-1]$ with $|{\cal E}|=h$. 
Set ${\cal E'}= {\cal E} \cup [ar+h:a(r+1)-1]$. Clearly, $|{\cal E'}|=a$. Hence, by erasure recovery property of $\mathcal{C}^*_{a,r}$ proved above, $\big\{c_i \mid i \in {\cal E} \cap [0:r-1] \big\}$ is recoverable using  $\big\{c_j \mid j \in [0:ar+h-1] \setminus {\cal E} \big\}$. From definition of $\mathcal{C}^*_{a,r}$, it can be seen that message symbols $\{c_j \mid j \in [hr:ar-1]\}$ are not involved in parity symbols $\{c_j \mid j \in [ar:ar+h-1]\}$. Hence, $\big\{c_i \mid i \in {\cal E} \cap [0:r-1] \big\}$ is recoverable using only $\big\{c_j \mid j \in \big([0:hr-1]\cup [ar:ar+h-1]\big) \setminus {\cal E} \big\}$. 
\eprf
\section{Extending to all parameters} \label{Sec:Ext}
In this section, we present rate-optimal $(a,\tau,r)$ LRSC for the case $\tau+1 \ne a(r+1)$. For $\tau+1 > a(r+1)$ we show that $\mathcal{C}_{(a,a(r+1)-1,r)}$ itself gives rate-optimal LRSC, whereas for $\tau+1 < a(r+1)$ a modified version of it works. 
\subsection{$\tau+1 > a(r+1)$}
It follows from definition  of LRSC that if $\tau+1>a(r+1)$, then any $(a,a(r+1)-1,r)$ LRSC is also an $(a,\tau,r)$ LRSC. Thus $\mathcal{C}_{(a,a(r+1)-1,r)}$ is a rate-optimal $(a,\tau,r)$ LRSC for all $\tau>a(r+1)-1$, since the rate of $\mathcal{C}_{(a,a(r+1)-1,r)}$ is $\frac{r}{r+1}$ which same as the rate upper bound \eqref{Eq:bound} for this case.
\subsection{$\tau+1 < a(r+1)$}
Note that if $\tau+1 < a(r+1)$, then $\min\left\{\frac{\tau+1-a}{\tau+1},\frac{r+1}{r}\right\}=\frac{\tau+1-a}{\tau+1}$. Hence our aim here is to construct an $(a,\tau,r)$ LRSC $\mathcal{C}_{(a,\tau,r)}$ of rate $\frac{\tau+1-a}{\tau+1}$ for all $\tau <a(r+1)-1$. Let $\tau+1-a=ur+v$, where $0 \le v < r$. Then $0 \le u <a$ as $\tau+1-a<ar$ and we set $\ell =a-u$. 
For this case, we fix $k=\tau+1-a$, $n=\tau+1$ and hence rate  $\frac{k}{n}=\frac{\tau+1-a}{\tau+1}$. For all $t \ge 0$ and $j \in [0:u-1]$, we set $(1 \times r)$ vector
\bean
\hat{\mu}_j(t)=\left[m_{jr}(t)~m_{jr+1}(t+1)~\dots~m_{jr+r-1}(t+r-1)\right].
\eean 
We also define $(1 \times r)$ vector  $\hat{\mu}_u(t)=\left[m_{ur}(t)~m_{ur+1}(t+1)~\dots~m_{ur+v-1}(t+v-1)~\mathbf{0}_{1 \times r-v}\right]$.\\

\begin{constr} Let $\tau+1<a(r+1)$.
	For all $t \ge 0$, the first $u$ parity symbols $\{p_i(t) \mid i \in [0:u-1]\}$ of $\mathcal{C}_{(a,\tau,r)}$ are defined as follows: 
	\bean 
	p_{i}(t)=\sum_{j=0}^{i}\hat{\mu}_{i-j}\left(t-r-j(r+1)\right) \underline{\Gamma}_{j}   
	+\sum_{j=i}^{u-1}\hat{\mu}_{u+i -j}\left(t-r-j(r+1)-v-\ell\right) \underline{\Gamma}_{a-u+j},
	\eean 
	for all $i \in [0:u-1]$.
	The remaining $\ell=a-u$ parity symbols take the form:
	\bean 
	p_{u+i}(t)&=& \sum_{j=0}^{u}\hat{\mu}_{u-j}\left(t-v-i-j(r+1)\right) \underline{\Gamma}_{j+i},
	\eean  
	for $i \in [0:a-u-1]$.
\end{constr}

\subsection*{Example:$(a=2,\tau=4,r=2)$ LRSC}

For this example, $k=3$, $n=5$ and $u=v=\ell=1$. We choose $C=\left[ \begin{array}{ccc}
1 & 1   \\
1 & 2   \\
\end{array}\right]$ over $\mathbb{F}_3$, resulting in $\Gamma =\left[ \begin{array}{ccc}
1 & 1   \\
1 & 2   \\ 
\end{array}\right]$. Then parity symbols of rate-optimal LRSC $\mathcal{C}_{(2,4,2)}$, as shown in Table~\ref{Tab:R12_}, are given by: 
\bean 
p_0(t)&=& \hat{\mu}_0(t-2)\underline{\Gamma}_0 + \hat{\mu}_1(t-4)\underline{\Gamma}_1 \\ &=& \left[ 
m_0(t-2)~m_1(t-1) \right]\left[ \begin{array}{c}
	1 \\ 1 
\end{array}\right] + \left[ 
m_2(t-4)~0\right]\left[ \begin{array}{c}
	1 \\ 2 
\end{array}\right] \\&=& m_0(t-2)+m_1(t-1)+m_2(t-4) ~~~\text{and} \\ p_1(t)&=& \hat{\mu}(t-1)\underline{\Gamma}_0 + \hat{\mu}_1(t-4)\underline{\Gamma}_1 \\ &=& \left[ 
m_2(t-1)~0\right]\left[ \begin{array}{c}
	1 \\ 1 
\end{array}\right]+\left[ 
m_0(t-4)~m_1(t-3) \right]\left[ \begin{array}{c}
	1 \\ 2 
\end{array}\right] \\ &=& m_0(t-4)+2m_1(t-3)+m_2(t-1). 
\eean 

\begin{thm}
	For any $(a, \tau, r)$ such that $a \le \tau$, $\tau+1 < a(r+1)$, ${\cal C}_{(a, \tau, r)}$ is an $(a, \tau, r)$ LRSC.
\end{thm}
\bprf
Assume that coded packet $\underline{c}(t)$ is erased and next $r$ coded packets are received. To recover message symbol $m_{xr+u}(t)$ for $x < u$ and $y < r$ which is an element in message vector $\hat{\mu}_{x}(t-y)$, we can use parity check:
\bean
p_x(t-y+r) = \sum\limits_{j=0}^x \hat{\mu}_{x-j}(t-y-j(r+1)) \underline{\Gamma}_j + \sum\limits_{j=i+1}^u \hat{\mu}_{u+i-j}(t-y-v-\ell-j(r+1))\underline{\Gamma}_{a-u+j}.
\eean
Notice that $\hat{\mu}_x(t-y)$ is the unknown vector in the RHS above as all other message vector have symbols from $\{\underline{m}(t'), t' < t \}$. The only unknown symbol in $\hat{\mu}_x(t-y)$ is $m_{xr+y}$ and hence it can be recovered. Similarly, for message symbols $m_{ur+y}(t)$ for $y < v$ that are elements in $\hat{\mu}_u(t-y)$ we can use parity check:
\bean
p_u(t-y+v) = \sum\limits_{j=0}^u \hat{\mu}_{u-j}(t-y-j(r+1)) \Gamma_j.
\eean
The only unknown element in the RHS above is $m_{ur+y}(t)$ and hence can be recovered. The parity checks used here for recovery have index $\le t+r$. Therefore ${\cal C}_{a, \tau, r}$ is an $(1, r)$ SC.


Now we show that $\mathcal{C}_{(a,\tau,r)}$ is an $(a,\tau)$ SC using  erasure recovery properties of $\mathcal{C}^*_{a,r}$  stated in Lemma~\ref{Lem:scalar}. Suppose we want to recover message packet $\underline{m}(t)$ in the presence of $a$ packet erasures in $[t:t+\tau]$, including $\underline{c}(t)$. We focus on recovery of symbol $m_{xr+y}(t)$ for some $x < u$, $y < r$ or $x = u$, $y < v$. Note that $m_{xr+y}(t)$ is a symbol in vector $\hat{\mu}_x(t-y)$. We look at the $a$ parity checks in which $\hat{\mu}_x(t-y)$ participates. 
The parity symbol $p_{i}(t+r+(i-x)(r+1)-y)$ for $x \le i \le u-1$ is given by:
\bean
p_{i}(t+r+(i-x)(r+1)-y)&=&\sum_{j=0}^{i}\hat{\mu}_{i-j}\left(t-y+(i-x-j)(r+1)\right) \underline{\Gamma}_{j} \\ &+&\sum_{j=i}^{u-1}\hat{\mu}_{u+i-j}\left(t-y+(i-x-j)(r+1)-v-\ell\right)\underline{\Gamma}_{a-u+j}.
\eean
Notice that we can compute $\hat{p}_{i-x}$ from $p_{i}(t+r+(i-x)(r+1)-y)$ as we know all message symbols $\underline{m}(t')$ with $t' < t$, where $\hat{p}_{i-x}$ is given by:
\bea
\label{eq:pc_short_uminusx}
\hat{p}_{i-x}=\sum_{j=0}^{i-x}\hat{\mu}_{i-j}\left(t-y+(i-x-j)(r+1)\right) \underline{\Gamma}_{j}, \ x \le i \le u-1.
\eea
The parity symbol $p_{u+i}(t-y+v+i+(u-x)(r+1))$ for $0 \le i \le a-u-1$ is given by:
\bean 
p_{u+i}(t-y+v+i+(u-x)(r+1)) = 
 \sum_{j=0}^{u}\hat{\mu}_{u-j}\left(t-y+(u-x-j)(r+1)\right) \underline{\Gamma}_{j+i}.
\eean  
Notice that we can compute $\hat{p}_{u-x+i}$ from $p_{i}(t-y+v+i+(u-x)(r+1))$ for all $i \in [0:a-u-1]$ as we know all message symbols $\underline{m}(t')$ with $t' < t$, where $\hat{p}_{u-x+i}$ is given by:
\bea
\label{eq:pc_short_aminusu}
\hat{p}_{u-x+i}=\sum_{j=0}^{u-x}\hat{\mu}_{u-j}\left(t-y+(u-x-j)(r+1)\right) \underline{\Gamma}_{j+i},\ 0 \le i \le a-u-1.
\eea

The parity check $p_{i}(t+r+(u+i-x)(r+1)-y+\ell+v)$ for $0 \le i \le x-1$ is given by:
\bean
p_{i}(t+r+(u+i-x)(r+1)+\ell+v-y)&=& \sum_{j=0}^{i}\hat{\mu}_{i-j}\left(t-y+\ell+v+(u+i-x-j)(r+1)\right) \underline{\Gamma}_{j} \\
&+&\sum_{j=i}^{u-1}\hat{\mu}_{u+i-j}\left(t-y+(u+i-x-j)(r+1)\right) \underline{\Gamma}_{a-u+j}.
\eean
Notice that we can compute $\hat{p}_{a-x+i}$ from $p_{i}(t+r+(u+i-x)(r+1)-y+\ell+v)$ as we know all message symbols $\underline{m}(t')$ with $t' < t$, where $\hat{p}_{a-x+i}$ is given by:
\bea
\nonumber \hat{p}_{a-x+i}&=&\sum_{j=0}^{i}\hat{\mu}_{i-j}\left(t-y+\ell+v+(u+i-x-j)(r+1)\right) \underline{\Gamma}_{j} \\
\label{eq:pc_short_x} &+&\sum_{j=i}^{u+i-x}\hat{\mu}_{u+i-j}\left(t-y+(u+i-x-j)(r+1)\right) \underline{\Gamma}_{a-u+j} \ 0 \le i \le x-1. 
\eea
From equations \eqref{eq:pc_short_uminusx}, \eqref{eq:pc_short_aminusu} and \eqref{eq:pc_short_x} we have that
\bean
\Big( \hat{\mu}_x(t-y), \ \hat{\mu}_{x+1}(t-y+(r+1)), \ \cdots, \ \hat{\mu}_{u-1}(t-y+(u-x-1)(r+1)), \\
\hat{\mu}_{u}(t-y+(u-x)(r+1)), \ \underbrace{\mathbf{0}, \ \cdots, \ \mathbf{0}}_{(a-u-1) \times r \text{ zeroes }}, \ \hat{\mu}_0(t-y+(u-x)(r+1)+v+\ell)\\ 
 \hat{\mu}_1(t-y+(u-x+1)(r+1)+v+\ell)), \cdots, \ \hat{\mu}_{x-1}(t-y+(u-1)(r+1)+v+\ell)\\
 \hat{p}_0, \cdots, \hat{p}_{a-1} \Big) \in {\cal C}_{a,r}^* \ \text{  defined in Lemma\ref{Lem:scalar}}.
\eean
Therefore by the property of ${\cal C}_{a,r}^*$ given in Lemma~\ref{Lem:scalar}, we can recover the erased symbols in $\mu_x(t-y)$ given there are at most $a$ erasures in the codeword. This is true as the codeword contains symbols coming from distinct packets with index $\le t+\tau$. Notice that it is true as $\hat{p}_{a-1}$ is obtained from $p_i(t-y+r+(u-1)(r+1)+\ell+v)=p_i(t-y+\tau)$. Thus we have argued that for all $\tau <a(r+1)-1$, $\mathcal{C}_{(a,\tau,r)}$ is a rate-optimal $(a,\tau,r)$ LRSC.
\eprf

\begin{note} For any $\tau >2 $,
	$\mathcal{C}_{(2,\tau,r)}$ with $r=\lceil \frac{\tau-1}{2}\rceil$ is an $(a=2,\tau)$ rate-optimal SC and can be constructed over any field of size $\ge r+1=\lceil \frac{\tau+1}{2}\rceil$.
\end{note}

\section{Conclusion and Future Work}
The notion of local recoverability in the context of streaming codes was introduced where apart from permitting recovery in the face of a specified number $a$ of erasures, the objective is to provide reduced decoding delay for the more commonly-occurring instance of a single erasure.  Rate-optimal constructions are provided for all parameter sets. The code also has the property of decoding delay that degrades gracefully with increasing number of erasures.  Our code construction requires large field size in general and field size reduction is left as future work.  Streaming codes ensuring packet recovery with decoding delay $\tau_1$ in the presence of $a_1$ erasures and delay $\tau_2$ for $a_2$ erasures also needs to be explored. 
Extending the idea of locality to streaming codes handling arbitrary and burst erasures is another interesting direction.   
\bibliographystyle{IEEEtran}
\bibliography{streaming}
\end{document}